\documentclass[conference]{IEEEtran}

\usepackage{graphicx}
\usepackage{balance}
\usepackage{comment}
\usepackage[noend]{algorithmic}
\usepackage{mathtools}
\usepackage[ruled,vlined]{algorithm2e}
\usepackage{amsmath}
\usepackage[caption=false]{subfig}
\usepackage{diagbox}
\usepackage{url}
\usepackage{float}
\usepackage{fancyhdr}

\DeclarePairedDelimiter{\ceil}{\lceil}{\rceil}

\DeclareMathOperator*{\argmin}{arg\,min}

\ifCLASSOPTIONcompsoc
  \usepackage[nocompress]{cite}
\else
  \usepackage{cite}
\fi
\ifCLASSINFOpdf
\else
\fi
\begin{document}
%
\title{\emph{FREE} - Fine-grained Scheduling for Reliable and Energy Efficient Data Collection in LoRaWAN}
%
%
%
%
\author{\IEEEauthorblockN{Khaled~Q.~Abdelfadeel\textsuperscript{1}, Dimitrios~Zorbas\textsuperscript{2}, Victor~Cionca\textsuperscript{3}, Dirk~Pesch\textsuperscript{4}}
\IEEEauthorblockA{\textsuperscript{1,}\textsuperscript{4} School of Computer Science and IT, University College Cork, Ireland\\
\textsuperscript{2}Tyndall National Institute, University College Cork, Ireland\\
\textsuperscript{3}Nimbus Research Centre, Cork Institute of Technology, Ireland\\
Corresponding Email: khaled.abdelfadeel@ieee.org}
}

\maketitle

\begin{abstract}

LoRaWAN promises to provide wide-area network access to low-cost devices that can operate for up to 10 years on a single $1000 mAh$ battery. This makes LoRaWAN particularly suited to data collection applications (e.g. monitoring applications), where device lifetime is a key performance metric. However, when supporting a large number of devices, LoRaWAN suffers from a scalability issue due to the high collision probability of its Aloha-based MAC layer. The performance worsens further when using acknowledged transmissions due to the duty cycle restriction at the gateway. For this, we propose \emph{FREE}, a fine-grained scheduling scheme for reliable and energy-efficient data collection in LoRaWAN. \emph{FREE} takes advantage of applications that do not have hard delay requirements on data delivery by supporting synchronized \textit{bulk data transmission}. This means data is \textit{buffered} for transmission in scheduled time slots instead of \textit{transmitted straight away}. \emph{FREE} allocates spreading factors, transmission powers, frequency channels, time slots, and schedules slots in frames for LoRaWAN end-devices. As a result, \emph{FREE} overcomes the scalability problem of LoRaWAN by eliminating collisions and grouping acknowledgments. We evaluate the performance of \emph{FREE} versus different legacy LoRaWAN configurations. The numerical results show that \emph{FREE} scales well and achieves almost $100\%$ data delivery and the device lifetime is estimated to over $10$ years independent of traffic type and network size. Comparing to poor scalability, low data delivery and device lifetime of fewer than $2$ years for acknowledged data traffic in the standard LoRaWAN configurations.

\end{abstract}



%
\IEEEpeerreviewmaketitle

\ifCLASSOPTIONcompsoc
\IEEEraisesectionheading{\section{Introduction}\label{introduction}}
\else
\section{Introduction}
\label{introduction}
\fi

%
%
%
%
LoRaWAN~\cite{adelantado2017understanding} has gained a lot of attention in the Internet of Things community recently due to its simple and open protocol stack as well as its deployment and management flexibility. LoRaWAN promises a \textit{deploy and forget} wireless sensing model and the high link budgets of its LoRa modulation~\cite{semtech2015lora} make LoRaWAN particularly suited to large-scale data collection applications~\cite{appsurvey} such as environmental monitoring applications. However, LoRaWAN experiences a high collision probability due to its Aloha-based MAC protocol. This results in a high packet loss rate, which affects the network reliability and scalability~\cite{bor2016lora}.

For example, in a common LoRaWAN configuration (spreading factor $12$, $125 kHz$ bandwidth) the data delivery rate decreases below $50\%$ for gateways serving more than $900$ devices~\cite{bor2016lora}. Even in terms of device lifetime, recent studies~\cite{energymodel} suggest a less than expected performance, e.g., a device lifetime of less than two years can be expected for a transmission interval of ten minutes, or five years for a $100$ minute interval. These results are for ideal situations without re-transmissions. The use of acknowledged transmissions to increase the delivery rate surprisingly does not help due to the duty cycle restriction at the gateway, which limits the number of acknowledgments that the gateway can send. This increases the traffic in the network due to the extra packets for re-transmissions, leading to even more collisions, which increases energy consumption and reduces device lifetime without improving the packet delivery ratio~\cite{pop2017does}. In summary, collecting data reliably over long periods of time from a medium-large number of battery-powered end-devices in LoRaWAN is still a challenge.

In this paper, we investigate a different approach to data collection in LoRaWAN. We propose to buffer the data at end devices and collect it during scheduled bulk transmissions at convenient points in time, instead of transmitting the data as soon as it is generated. The advantage of this approach is that we can schedule transmissions in an efficient way, thus extending the battery lifetime of the devices. We show that this approach can achieve \textit{(a)} a more than five-fold increase in device lifetime, passing the ten year mark, and \textit{(b)} a $>99\%$ data delivery ratio. The cause of the improvements is due to scheduled transmissions that eliminate collisions and the use of longer packets that reduce the overhead of MAC headers. We call our proposed approach \textit{FREE} - Fine-grained scheduling for Reliable and Energy Efficient data collection. \emph{FREE} is a bulk data collection protocol for LoRaWAN that can support a wide range of delay-tolerant applications such as smart city applications (e.g., vehicle and railway infrastructure monitoring~\cite{hodge2014wireless}, traffic monitoring~\cite{nellore2016survey}, air quality monitoring~\cite{airquality}, smart metering~\cite{smartmetering} etc.), and remote sensing applications (e.g. volcanoes~\cite{reventador}, glaciers~\cite{permafrost}, precision agriculture~\cite{baggio2005wireless}, etc.). Another suitable use case for \emph{FREE} is the data mule~\cite{datamuling}, a discontinuous data collection mechanism for large remote geographical areas without wired network connectivity. Here, a LoRaWAN gateway (e.g., carried by a mobile vehicle~\cite{zorbas2018collision}) is periodically brought into the communication range of the end devices for data collection. Despite its high degree of practicality, to the best of our knowledge, a bulk data transmission approach for LoRaWAN has not yet been reported in the literature.

\emph{FREE} replaces the standard simple random access method used by LoRaWAN with a coordinated medium access approach. Specifically, network devices are synchronized and transmissions are scheduled. Although such coordinated access has been investigated before in the general context of wireless networks, our proposal takes a new twist due to LoRaWAN's unique characteristics. LoRaWAN transmissions are restricted by the regulatory duty cycle limitations~\cite{etsi}, which is a key constraint for networks operating in unlicensed bands and do not adopt a \textit{listen-before-talk} policy. In addition to that LoRaWAN gateways can decode multiple concurrent transmissions (i.e. up to $8$) as long as they use different channels and/or different spreading factors; although inter-spreading factor interference may appear due to the imperfect orthogonality of LoRa~\cite{croce2018imperfect}. Additionally, downlink transmissions can only take place after uplink transmissions~\cite{adelantado2017understanding}\footnote{In this work, we consider the most limited LoRaWAN devices (i.e. Class-A devices)}. These unique characteristics impose challenges on computing the schedule and maintaining the synchronization, which have not been studied before in time coordinated networks~\cite{demirkol2006mac} and, particularly, for bulk data transmission~\cite{liang2008koala}.

The scheduling algorithm in \emph{FREE} is executed centrally at the gateway based on information collected from end devices, e.g. the amount of buffered data and the actual path loss. The schedule maximizes the throughput by allocating spreading factors, channels, and transmission powers to devices so that concurrent transmissions can be successfully decoded at the gateway. Our schedule design utilizes six parallel frames, one per spreading factor, and some of the frames use multiple channels. The channels and transmission powers are allocated such that the impact of imperfect spreading factor orthogonality is minimal. The scheduling algorithm runs in a \textit{greedy online} fashion with the objective to minimize data collection time and energy consumption while obeying duty cycle regulations. Synchronization, which is crucial for running effectively the schedule, is performed in two stages. The first stage is used for coarse synchronization up to $1sec$ accuracy and the second stage is used for finer synchronization up to $1ms$ accuracy. \emph{FREE} supports unconfirmable (unacknowledged) and confirmable (acknowledged) uplink data traffic, where the latter one uses a compressed group acknowledgment scheme to work around the duty cycle limitations at the gateway.

The next section describes the problem in more detail and the \emph{FREE} synchronization mechanism. The design of the \emph{FREE} scheduling approach is presented in \S\ref{free}, including the required allocation algorithms. \S\ref{evaluation} provides a performance evaluation. The related work section is provided in \S\ref{relatedwork} and finally the paper is concluded in \S\ref{conclusion}.

 
 
\begin{figure}[!t]
  \centering
  \includegraphics[width=1\columnwidth]{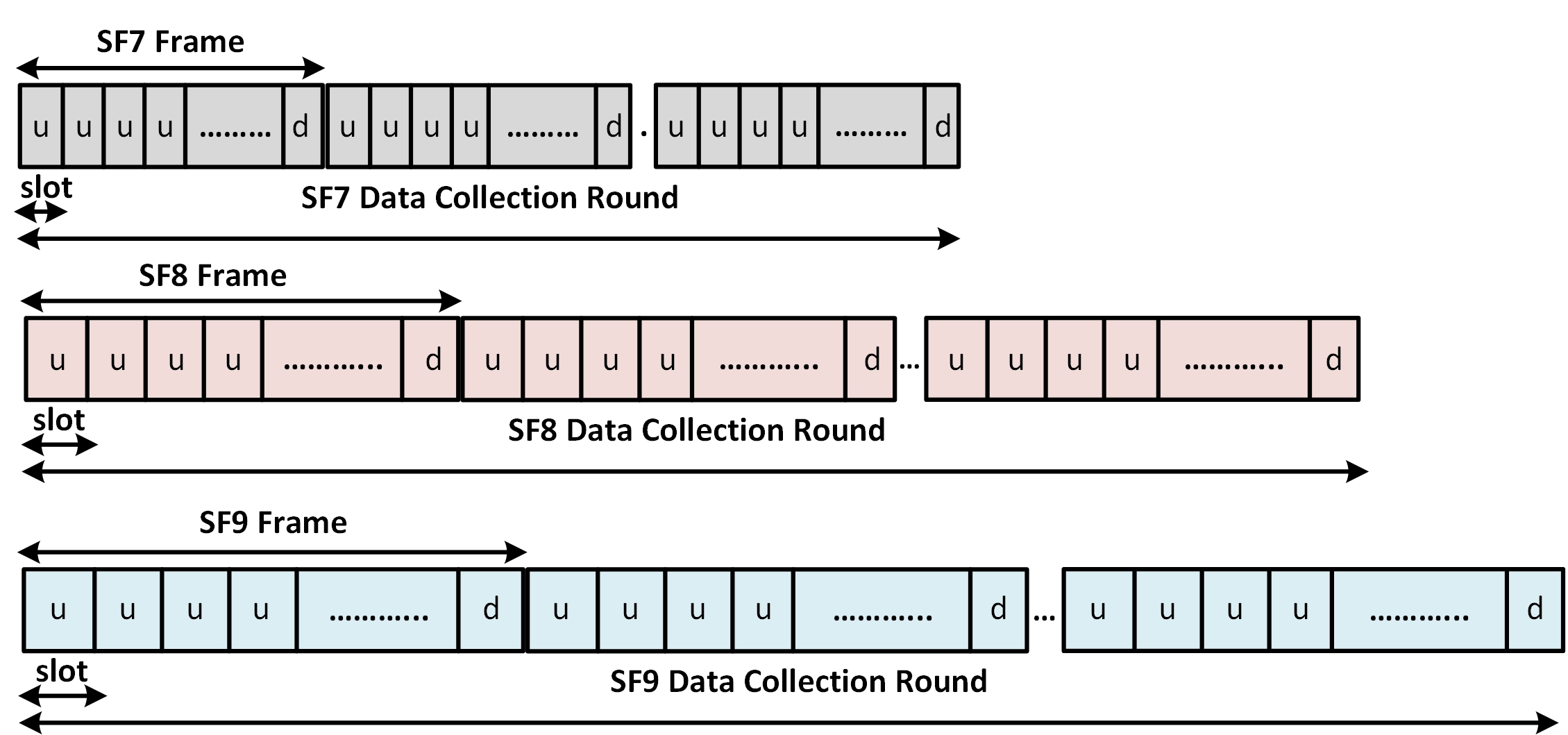}
  \caption{\emph{FREE} Frame Structures (u: Uplink and d: Downlink)}\label{fig:framestructure}
  \vspace{-0.35cm}
\end{figure}

\section{System Description} \label{sysmodel}

We consider a private LoRaWAN deployment of one gateway and a number of end devices without interference from other LoRaWAN deployments or other technologies operating in the same frequency band. The devices' applications are sensing or data collection applications, where data is not time critical for a period of time. These applications might generate the data either periodically or event based or in a mixture of both. In this scenario, the devices can buffer the data in local storage and send it later in bulk. This description fits well a lot of applications in the domain of LoRaWAN such as monitoring applications in smart cities~\cite{airquality,smartmetering} and remote areas~\cite{reventador,permafrost}. Also, it fits the use cases in which the gateway is temporarily accessible due to either duty-cycling to save power (e.g. battery-powered) or moving (e.g. data mule~\cite{zorbas2018collision}).

Once data collection starts, the devices have to transmit their buffered data to the gateway. The amount of this data could be large, depending on the data generation rate and the collection periodicity. With an Aloha-based MAC, transmissions would suffer from severe collisions, leading to high energy consumption and prolonged collection time. This is due to that the medium would be almost saturated and, as known, asynchronous access like Aloha does not perform well in such a scenario. To alleviate the impact of collisions, transmissions have to be scheduled based on synchronous access.

\subsection{Proposed Solution - \emph{FREE}}

\emph{FREE} is a fine-grained scheduling approach to synchronize transmissions to achieve reliable and energy-efficient data collection. Data collection takes place at periodic/aperiodic intervals known to the gateway and end devices. In order to compute the schedule of each data collection period, the gateway needs to know the number of end devices that have data to transmit, the amount of their buffered data, and estimate their path loss. The gateway then disseminates the schedule and synchronizes all end devices with the same time reference. Computing and disseminating the schedule and maintaining the synchronization happen in two stages as will be illustrated in \S~\ref{synchronizationphase}.

\emph{FREE} supports three channels of $1\%$ duty cycle as in LoRaWAN plus a channel of $10\%$ duty cycle in the EU $868$ ISM band. The duty cycle defines the maximum time and the maximum percentage with which a device can occupy a channel. For instance, $1\%$ duty cycle results in a maximum transmission time of $36 s/h$ for each device. Additionally, a minimum silent time of $99 T$, where $T$ is the packet transmission time, is required after each transmission. The three $1\%$ channels are used for uplink and downlink and the $10\%$ channel is only dedicated for downlink traffic. Before \emph{FREE} has constructed a schedule, the channels are accessed in an asynchronous manner, but once the schedule is calculated and disseminated to end devices, channel access takes place in a synchronous manner.

The transmissions in \emph{FREE} are scheduled in such a way that devices with the same spreading factor transmit sequentially and devices with different spreading factors transmit simultaneously. For this purpose, time is divided into frames, with each frame having a number of \textit{uplink} slots and one \textit{downlink} slot at the end of a frame as depicted in Fig.~\ref{fig:framestructure}. There are six \textit{parallel} frames, corresponding to LoRa's six spreading factors ($7-12$). Devices per spreading factor are assigned separate slots in the corresponding frame, where the schedule remains the same in the consecutive frames. The term \textit{round} per spreading factor is used to indicate the consecutive frames that are required to collect all data. In addition to the parallel rounds, \emph{FREE} may also support multiple channels per spreading factor. The round, in this case, starts with one slot delay on the second channel, two slots on the third channel and so on. This is because devices cannot transmit simultaneously on multiple channels. Devices (using this particular spreading factor) transmit in each frame on all supported channels, which speeds up the corresponding data collection round. The slot and the frame lengths are equal for the same spreading factor but may not be equal for different spreading factors. The slot length and the number of uplink slots per frame depend on multiple factors, e.g. packet length, number of devices per corresponding spreading factor, etc. All factors that affect \emph{FREE}'s design are examined in \S~\ref{free}.

\begin{figure}[!t]
  \centering
  \includegraphics[width=1\columnwidth]{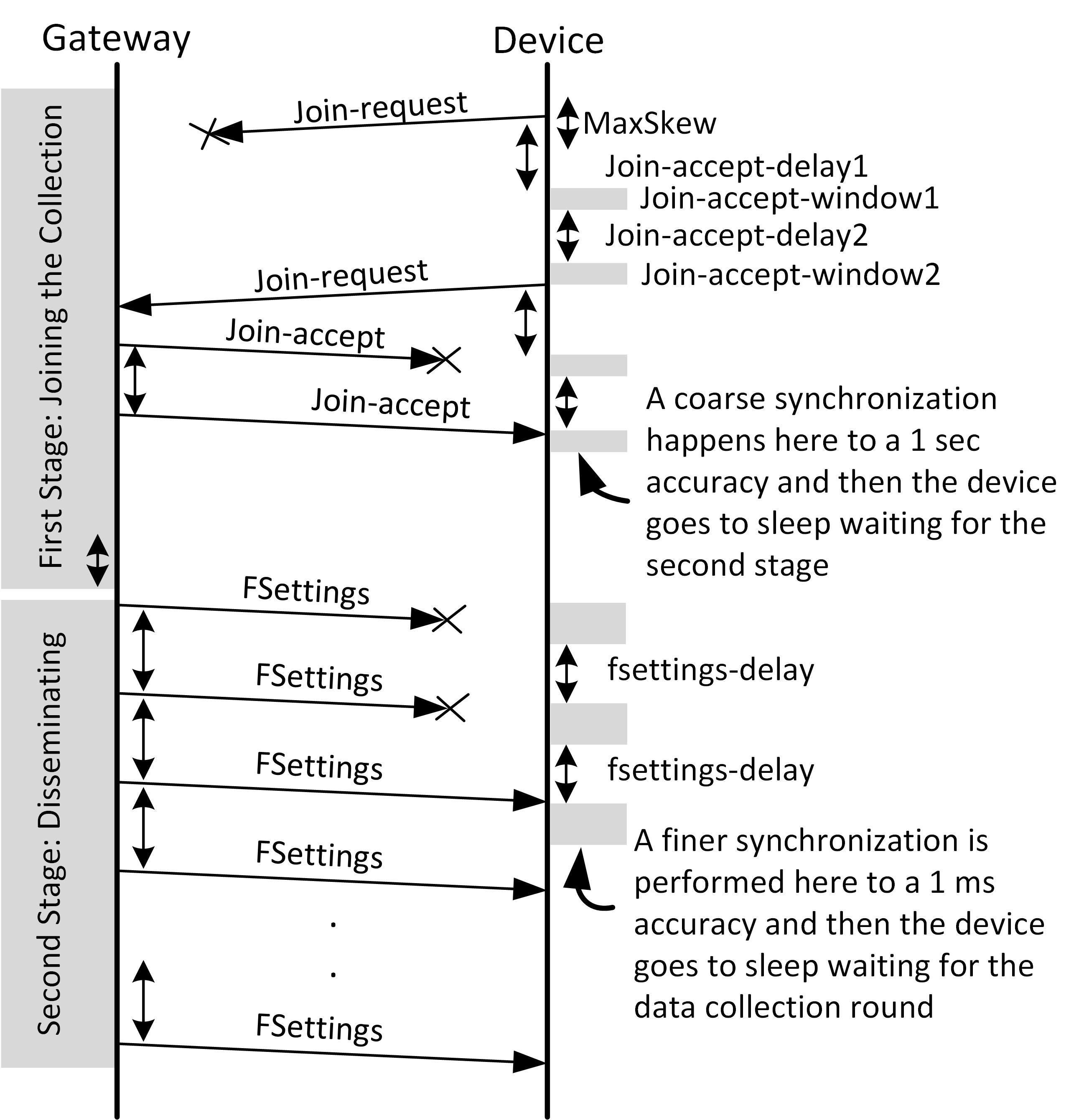}
  \caption{The Sequence of Actions in the Joining and Synchronization Phase}\label{joiningandsynchronization}
  \vspace{-0.35cm}
\end{figure}

Before starting the data collection, the network entities (i.e. the gateway and end devices) need to agree on a unified schedule and a time reference in order to synchronize transmissions. These agreements are performed in what is called ``the joining and synchronization phase'' as depicted in Fig.~\ref{joiningandsynchronization}. The next subsection describes the joining and synchronization phase in detail.

\subsection{Joining and Synchronization Phase} \label{synchronizationphase}

This phase consists of two consecutive stages as shown in Fig.~\ref{joiningandsynchronization}. The structure of all messages used in Fig.~\ref{joiningandsynchronization} is illustrated in appendix~\ref{annexA}. The purpose of the first stage is to collect schedule requests from end devices and provide them with their transmission parameters. In addition to that the first time synchronization happens to a $1 sec$ accuracy. The second stage uses the gathered information from the first stage to compute and broadcast the schedule for all devices. This stage also increases the time synchronization to less than $1 msec$. We assume here that the duration and the channel(s) of both stages are globally shared among devices in advance.

\subsubsection{First Stage: Joining the Data Collection}

Each device, if it has data to send, submits the intention to transmit in the next data collection round by sending a \textit{join-request} message to the gateway, using the lowest possible spreading factor~\cite{abdelfadeel2018fadr}. In this message, devices reveal the amount of their buffered data \textit{DataSize} and their delay elasticity \textit{DElasticity} in seconds. Once the gateway receives a \textit{join-request} message, it estimates the path loss of the corresponding device and responds with a \textit{join-accept} message. A \textit{join-accept} message guides a particular device to set up its data rate \textit{DataRate}, transmission power \textit{TxPower}, channel(s) \textit{ChMask}, slot number in the assigned frame \textit{SlotFrame}, and the remaining time in seconds to the starting time of the second stage \textit{Secondstage}. Using \textit{SecondStage}, devices can synchronize themselves to be less than one second out of synchronization with reference to the gateway. Once a device has received the \textit{join-accept} message, it goes into sleep mode waiting for the second stage.

The \textit{join-requests} are transmitted in an asynchronous fashion using LoRaWAN's default Aloha MAC. In this case, each device keeps sending \textit{join-request} messages until it receives a \textit{join-accept} message from the gateway or exceeds the maximum duration of this stage. These requests are transmitted on the three ($1\%$ duty cycle) channels, where a device randomly chooses one of the channels to send its requests. The gateway, if its duty cycle permits, sends the \textit{join-accept} message on the same channel as the \textit{join-request} during the first receive window or on the extra downlink ($10\%$ duty cycle) channel during the second receive window.



\subsubsection{Second Stage: Disseminating Frame Structures}

The gateway broadcasts the \textit{FSettings} message, encoded using the highest spreading factor, during this stage in order to reach all devices. 
The \textit{FSettings} message contains the packet lengths \textit{PcktSizes}, guard times \textit{Guards} in milliseconds, and the numbers of slots per frames \textit{FrameLens} for all spreading factors. 
Using the corresponding \textit{PcktSize} and \textit{Guard} values, each device calculates the slot length using $T_{PcktSize} + 2 Guard\,\,\,\,[ms]$, where $T_{PcktSize}$ is the transmission time as defined in \cite{lora2013calculator}. Additionally, the \textit{FSettings} message communicates the remaining time in milliseconds to the starting time of the data collection round \textit{DataCollection} and the remaining time in seconds to the next joining and synchronization phase \textit{NextRound}. Once a device receives one of the \textit{FSettings} messages, it goes into sleep mode waiting for the data collection rounds to start.

\subsection{Data Collection Rounds} \label{dcr}

In order to transmit its buffered data, each device wakes up for the right slot, 
waits for one guard time before transmitting a packet 
and then goes back to sleep mode. Using this method, the transmission takes place within the assigned slot and no overlapping with other slots occurs even in case of de-synchronization among devices. 
As long as a device has data to send, it will wake up for the same slot in subsequent frames and will repeat the same procedure.

In case of a confirmable uplink transmission, the device wakes up again before the last slot in the frame to receive the acknowledgment message. If a device could not receive the acknowledgment message due to a transmission error, the device will re-transmit the same packet up to 8 times as specified in the LoRaWAN standard, before dropping it. The gateway acknowledges all received packets during the frame at once using a bitmap message. The bit position in this message corresponds to the slot number in the frame. 
A bit equalling $1$ in the acknowledgment message indicates a correct packet reception in the corresponding slot, otherwise nothing or a corrupt packet was received.

The data collection periodicity (\textit{NextRound}) is subject to the delay elasticity of the running applications (\textit{DElasticity}). Specifically, the data collection has to be performed more often than the lowest delay requirements. For instance, if the network hosts two applications, where the first has a delay elasticity of $12-hours$ and the second one of $24-hours$, then the data collection will be performed every $12-hours$ or less to satisfy the delay requirements of both applications.

\begin{table}[!t]
  \centering
  \caption{Notations used in this paper} \label{notations}
  \begin{tabular}{c | l}
  \textbf{Symbols} & \textbf{Notations} \\
  \hline
  \hline
  $f\in\mathcal{F}, b\in\mathcal{B}, c\in\mathcal{C}$ & Spreading factor, bandwidth, coding rate \\
  $P_{f,\mathrm{BER}}, P_{f,\mathrm{PER}}$ & Bit error rate, Packet error rate \\
  $L_f, H$ & Packet length per $f$, MAC header length \\
  $\omega_f, SNR_f$ & Receiver sensitivity, SNR limit per $f$ \\
  $T_f$ & Transmission time of $L_f+H$ packet per $f$ \\
  $M_f$ & Number of assigned channels per $f$ \\
  $d$ & Duty cycle \\
  $N_f$ & Number of assigned devices per $f$\\
  $G_f$ & Guard time per $f$ \\
  $S_f$ & Number of slots in the frame per $f$ \\
  \end{tabular}
\end{table}

\section{Fine-tuning a \emph{FREE} schedule} \label{free}

The \emph{FREE} approach schedules LoRaWAN transmissions in a manner so as to speed up the overall data collection and minimize the overall energy consumption. However, LoRa transmissions are affected by many factors that are technology-dependent, e.g., duty-cycle, multiple spreading factors, etc. These factors result in interesting trade-offs in the schedule design. In the next subsections, these trade-offs are presented and allocation algorithms for the transmission parameters are proposed. The notations used throughout this paper are summarized in Table~\ref{notations}.

\subsection{Packet Lengths} \label{pcktlen}

The packet length (i.e. a chunk of devices' buffered data) is both directly and inversely proportional to the energy consumption. On one hand, longer packets have a higher probability of error, which then requires re-transmissions for successful delivery and, therefore, higher energy consumption. On the other hand, if shorter packets are used, the ratio of payload to MAC header is reduced, so a higher number of transmissions are needed to send a certain amount of data, resulting in higher energy consumption. In this section, we address this trade-off theoretically to compute the packet length that minimizes the overall energy consumption.

\begin{table}[!t]
  \centering
  \caption{$SNR$ minimum limits [$dB$]} \label{snrlimts}
  \begin{tabular}{c||c|c|c|c|c|c}
  \textbf{$f$} & 7 & 8 & 9 & 10 & 11 & 12 \\
  \hline
  \textbf{$SNR_f$} & $-6$ & $-9$ & $-12$ & $-15$ & $-17.5$ & $-20$ \\
  \end{tabular}
\end{table}

We first need the Bit Error Rate (BER) of the LoRa modulation for a given spreading factor $f$ as per Eq.~\ref{ber}~\cite{reynders2016range}.
\begin{align}
P_{f,\mathrm{BER}} =&\mathcal{Q}(\frac{\log_{12}f}{\sqrt(2)}\frac{E_{b}}{N_{0}}),  \label{ber}
\end{align} where $E_{b}/N_{0}$ denotes the energy per bit to noise power spectral density ratio and $\mathcal{Q}(x)$ is the Q-function. $E_b/N_0$ can be converted into Signal-to-Noise-Ratio (SNR) as follows~\cite{understanding2012}:
\begin{align}\label{snrtoepb}
E_{b}/N_{0} = &SNR_f -10\log{\frac{b}{2^{f}}} - 10\log{f} \nonumber \\
&- 10\log{c} +10\log{b},
\end{align} where $b$ and $c$ are the bandwidth and coding rate respectively, and $SNR_f$ is the SNR limit per spreading factor $f$. These limits are given in Table~\ref{snrlimts} as reported in ~\cite{lora2013calculator}.

Using Eq.~\ref{ber}, the packet error rate can be written as:
\begin{align}
P_{f,\mathrm{PER}} =&1- (1- P_{f,\mathrm{BER}})^{8(L_f+H)}. \label{per}
\end{align} Eq.~\ref{per} assumes independently distributed and constant bit error across a packet, considering a packet to be corrupted if one or more of its bits is corrupted. Although this assumption may not always hold in reality, it is reasonable approach as it yields the worst case packet error rate. In Eq.~\ref{per}, $L_f$ is the payload length of a packet using spreading factor $f$ and $H$ is the length of the MAC header. Given Eq.~\ref{per}, the expected number of re-transmissions can be written as ~\cite{de2005high} 
\begin{align}
R_f =&\sum_{n=1}^{\infty} P^{n}_{f,\mathrm{PER}} = \frac{P_{f,\mathrm{PER}}}{1-P_{f,\mathrm{PER}}}. \label{retrans}
\end{align} In this case, the energy consumption of transmitting the total buffered data of size $S$ can be calculated using Eq.~\ref{energy}. 
\begin{align}
E_f =&(1+R_f) \ceil{\frac{S}{L_f}} T_{f} I V, \label{energy}
\end{align} where $T_{f}$ is the transmission time for sending a packet of length $L_f+H$, and $I$ and $V$ are average current and voltage in the transceiver chip during transmission, respectively.

\begin{figure}[!t]
  \centering
	\includegraphics[width=0.7\columnwidth]{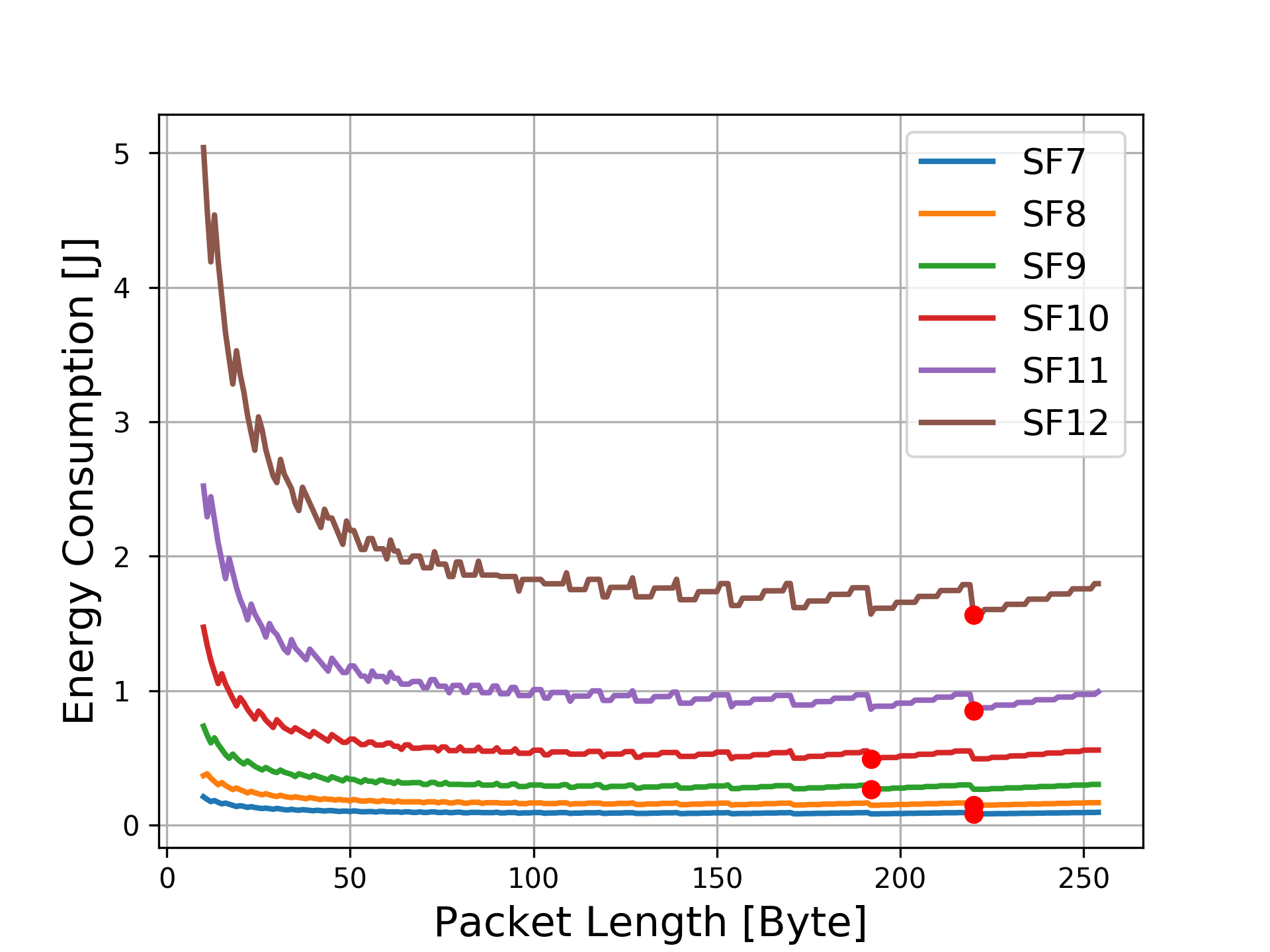}
  \caption{Energy Consumption at different Packet Lengths}\label{fig:energy}
  \vspace{-0.35cm}
\end{figure}

Figure~\ref{fig:energy} shows an example of the energy consumption in the case of transmitting buffered data of size $1.5 KB$. We examined all packet lengths, starting from $20 B$ length to $255 B$, which is the longest packet length permitted in LoRa. The point on each curve represents the packet length that achieves the minimum energy consumption for the corresponding spreading factor. As shown, the packet length for all spreading factors tends to be large to minimize energy. The conclusion from this evaluation is that the impact of packet errors is not as critical as the impact of the MAC header overhead in terms of energy consumption. For this reason, long packets for all spreading factors are better than short packets to reduce the overall number of transmissions and, thus, the impact of MAC headers.

\subsection{Spreading Factor and Slot Number} \label{sfallocation}

The spreading factor allocation affects the overall network performance in terms of energy consumption and data collection time. First, the spreading factor allocation determines how many slots are in each frame, which affects the time of the data collection rounds and, thus, the overall data collection time. Second, the spreading factor determines the transmission time of packets and, thus, determines the energy consumption \cite{abdelfadeel2018fadr}. Indeed, the energy consumption increases with higher spreading factors. In this section, we investigate the spreading factor allocation along with our objectives for \emph{FREE} to minimize data collection time and energy consumption.

Generally speaking, the spreading factor allocation is constrained by the specific path loss between a device and its serving gateway as not every spreading factor can be used for any device~\cite{adelantado2017understanding}. This is due to spreading factors having different receiver sensitivities, which can be calculated as follows:
\begin{align}\label{sensitivity}
\omega_f = -174 + 10\log{b} + NF + SNR_f.
\end{align}
The first term denotes the thermal noise per 1 Hz bandwidth and $NF$ denotes the receiver noise figure. The receiver sensitivity will influence the spreading factor allocation as a device must have a higher Received Signal Strength Indicator (RSSI) than the receiver sensitivity associated with the corresponding spreading factor. The standard LoRaWAN spreading factor allocation algorithm meets the receiver sensitivity constraint by usually assigning the lowest possible spreading factor. Although this approach could minimize the energy consumption, it might lead to excessive use of certain spreading factors. This can lead to an unbalanced schedule that is sub-optimal in the data collection time. Therefore, the objectives of minimizing energy consumption and collection time lead to contradictions in terms of the spreading factor allocation. In order to address this, we propose Algorithm~\ref{sfah} with two different objective functions, which we trigger with the $\alpha$ flag, e.g. $\alpha \in\{0,1\}$. When $\alpha=1$, the algorithm allocates spreading factors so as to minimize the data collection time, whereas, the energy minimization objective is considered when $\alpha=0$.

\begin{algorithm}[!t]
\LinesNumbered
\SetKwInOut{Input}{input}
\SetKwInOut{Output}{output}
\Input{$\alpha\in\{0,1\}$,$\mathcal{F}$, $\omega_{f}\forall f\in\mathcal{F}$, $T_f \forall f\in\mathcal{F}$, $M_{f}\forall f\in\mathcal{F}$, $H$, $d$, $I$,$V$}
\lForEach{$f \in \mathcal{F}$}{$N_f = \emptyset$}
\While{Stage 1}{
\If{Received a Join-request}{
\Input{$rssi$, $DataSize$}
\Output{Spreading Factor, Slot number}
$minf = \argmin_{f \in \mathcal{F}} rssi > \omega_{f}$\;
$tempf = \argmin_{f \in \mathcal{F}} cost$
\If{$f >= minf$}{
\begin{align}
cost =
\smash[t]{\begin{cases}
&\Big[\max(X_{f}+1,\ceil{1/d})\ceil{\\
&\frac{DataSize}{L_{f}M_{f}}}+(M_{f}-1)\Big] T_f  \text{\textbf{ if }} \alpha == 1\\[3pt]
&\ceil{\frac{DataSize}{L_{f}}} T_f I V \text{\textbf{ if }} \alpha == 0\nonumber
\end{cases}}
\end{align}
}
$X_{tempf}$ += 1\; 
\Return{$tempf, X_{tempf}$}\;
}
}
\caption{$\alpha$-Spreading Factor Allocation and Slots}\label{sfah}
\end{algorithm}

Once the gateway receives a \textit{join-request} message from a device, it extracts the amount of data the device wants to send as well as the RSSI of the message. Then, Algorithm~\ref{sfah} runs to determine the spreading factor that the device should use according to the objective of the allocation. Algorithm~\ref{sfah} checks, firstly, the minimum spreading factor that can be assigned based on the minimum sensitivities (line 4). Then, it evaluates the cost function for all \textit{allowed} spreading factors (i.e. equal or higher than the minimum spreading factor) and assigns the spreading factor that achieves the minimum cost. Subsequently, the next available slot in the corresponding frame is allocated to the device (line 8).

\subsubsection{Minimum Energy Consumption ($\alpha=0$)}

In the case of minimizing the energy consumption, the following cost function is used to determine the best spreading factor:
\begin{align}\label{cost2}
Cost_{f}=\ceil{\frac{DataSize}{L_{f}}} T_f I V \,\,\,\,\,\,\,\,\,\text{with} \,f\,\in\,\mathcal{F},
\end{align}
This cost function calculates roughly the energy consumption using spreading factor $f$. The $I$ and $V$ are average current and voltage in the transceiver chip during transmission, respectively.

\subsubsection{Minimum Collection Time ($\alpha=1$)}

Devices with the same spreading factor are assigned to consecutive slots in a frame, which determines the frame length. In order to minimize the data collection time, balancing the frame lengths over all spreading factors is required to take as much advantage of concurrent transmissions as possible. Therefore, in some network deployments, it is better to switch certain devices to higher spreading factors than what might be optimal from the energy consumption perspective.

However, each frame has a minimum length that cannot be lowered even if there is only one device using the corresponding spreading factor. The minimum length depends on the corresponding packet length and the duty cycle applied. A minimum frame length is required to allow a device to reuse the same slot number in the following frames without violating the duty cycle. Given the set of spreading factors $\mathcal{F} = \{7,.,f,.,12\}$ and the duty cycle $d\in(0,1]$, the minimum frame lengths can be written as $T_f/d\,\text{with}f\in\mathcal{F}$. In case of the slot length equaling $T_f$, the minimum number of slots in a frame should be equal to $1/d$.

In this case, the optimal spreading factor for the device is the one that minimizes the following cost function:
\begin{align}\label{cost1}
Cost_{f} = &\Big[\max(X_{f}+1,\ceil{1/d}) \ceil{\frac{DataSize}{L_{f}M_{f}}} \nonumber \\
           &+(M_{f}-1)\Big] T_f \,\,\,\,\,\,\,\,\,\text{with} \,f\,\in\,\mathcal{F},
\end{align} where $X_{f}$ is the number of devices that already have been assigned with spreading factor $f$ and $M_f$ is the number of channels per spreading factor $f$. The above cost function calculates roughly the data collection time for this device based on the previous knowledge of $X_{f}$. $M_{f}-1$ represents the required extra slots in case more than one channel is available for spreading factor $f$ because the frame starts with one slot delay in each channel compared to the previous channel.

\subsubsection{Allocation Optimality}

The spreading factor allocation is performed in the first stage. Through this stage, the gateway has only partial information about the network, i.e., received \textit{join-requests} so far. Therefore, the allocation algorithm is performed in a \emph{greedy online} manner based on this partial available knowledge. The online fashion of the allocation is crucial as it makes our approach \textit{independent} of the network topology and application(s) (i.e. event-based or periodic). This is because the allocation algorithm does not constrain the number of devices that can participate in each data collection, nor the amount of their buffered data or their positions from the gateway. This is important, in particular, for the data mule use case, where the gateway might connect to the LoRaWAN deployment from different positions or in a use case, where end devices may move between data collections. Taking into account this independence, \emph{FREE} allocates the \textit{best} possible spreading factor for each device, according to the chosen objective function. In~\cite{offline2019}, we investigated an \textit{offline} version of the spreading factor allocation in which the global state information of the network is assumed to be known in advance of the data collections. The offline algorithm works by scheduling transmissions in a decreasing order of the spreading factors. Although it achieves shorter data collection time, the improvement over \emph{FREE} is not significant. This is because the \textit{join-requests} from devices with low minimum spreading factor are more likely to be received before those with high minimum spreading factors, which gives the allocation in FREE more flexibility. This results in the schedule in \emph{FREE} to be comparable to the schedule in~\cite{offline2019} even without knowing the global state information of the network.

\subsection{Transmission Power and Channels}

\begin{table}[!t]
  \centering
  \caption{Interference Thresholds [$dBm$] \cite{croce2018imperfect}} \label{isolationthresholds}
\begin{tabular}{c||*6c}
    \backslashbox{\textbf{Ref.}}{\textbf{Int.}} & \textbf{$SF_7$} & \textbf{$SF_8$} & \textbf{$SF_9$} & \textbf{$SF_{10}$} & \textbf{$SF_{11}$} & \textbf{$SF_{12}$}\\
    \hline
    \hline
    \textbf{$SF_7$} & $1$ & $-8$ & $-9$ & $-9$ & $-9$ & $-9$\\
	\textbf{$SF_8$} & $-11$ & $1$ & $-11$ & $-12$ & $-13$ & $-13$\\
	\textbf{$SF_9$} & $-15$ & $-13$ & $1$ & $-13$ & $-14$ & $-15$\\
	\textbf{$SF_{10}$} & $-19$ & $-18$ & $-17$ & $1$ & $-17$ & $-18$\\
	\textbf{$SF_{11}$} & $-22$ & $-22$ & $-21$ & $-20$ & $1$ & $-20$\\
	\textbf{$SF_{12}$} & $-25$ & $-25$ & $-25$ & $-24$ & $-23$ & $1$\\
    \end{tabular}
\end{table}

\emph{FREE} exploits concurrent LoRa/LoRaWAN transmissions due to different spreading factors. However, the error rate of concurrently received transmissions is affected by the differing strengths of the received signals. This is because the spreading factors are not fully orthogonal and, thus, cause mutual interference. The orthogonality property of spreading factors has been experimentally studied in~\cite{croce2018imperfect}. Table~\ref{isolationthresholds} presents the interference thresholds among spreading factors. If two signals (using different spreading factors on the same channel) overlap and the difference between their signal strengths is less than the corresponding interference threshold, the two signals can be successfully received. Otherwise, only the strongest signal can be detected. Therefore, careful isolation is required in order to receive all the concurrent transmissions successfully. As is usual in spread spectrum communications, the isolation can be performed in power (i.e. control transmission power) or in frequency (i.e. assigned channels).

Controlling the transmission power to minimize the cross spreading factor interference would greatly complicate the scheduling. This may lead to different transmission powers per device per transmission, which would require sending frequent power control commands to devices. We believe that complicated transmission power control is not required in \emph{FREE} as the number of concurrent transmissions on the same channel is controllable. For that reason, we rely less on varying the transmission power and depend more on the channel assignment. There are two constraints that are governing the channel assignment. Firstly, the limited number of uplink channels and, secondly, the maximum number of concurrent transmissions that the gateway can handle. Increasing the number of channels per spreading factor expedites the data collection by enabling more concurrent transmissions. However, this may lead to more transmission rejections by the gateway if the number of concurrent transmissions exceeds the maximum capacity of the gateway. In this work, similar to standard LoRaWAN, we consider a maximum of \textit{three} uplink channels and a maximum number of \textit{eight} concurrent transmissions, which is supported by most LoRaWAN gateways.

Therefore, \emph{FREE} allocates one channel to each spreading factor from 7 to 10 and two channels to spreading factors 11-12. Using this allocation, the gateway receives eight transmissions on all channels at a maximum. The reason behind allocating two channels to spreading factors 11 and 12 is to reduce their data collection rounds by about 50\%. As their minimum frame lengths are large, the overall collection time would be negatively affected even if only a small number of devices used these particular spreading factors. However, this leaves the question as to what is the criteria for assigning the spreading factor against the channels?

In \cite{abdelfadeel2018fair}, we found that the majority of cross spreading factor collisions are favoring low spreading factors, particularly 7 over the high spreading factors due to their high interference threshold (see Table~\ref{isolationthresholds}) compared to the other spreading factors. Therefore, \emph{FREE} assigns a separate channel to devices that use spreading factor 7 (e.g. channel\#1). The assignment of channel and transmission power is shown in Table~\ref{txandchanl}. Additionally, \emph{FREE} combines the transmissions of spreading factors 9 and 10 with one of the transmissions of 11 and 12 onto another separate channel (e.g. channel\#2). The remaining transmissions are assigned the last channel (e.g. channel\#3). The reason behind separating spreading factor 8 from 9 and 10 is also the high interference threshold of 8 over 9 and 10.

The frequency isolation of spreading factor 7 allows its transmissions to use the maximum transmission level (i.e. $14dBm$), which increases the link reliability without adding any cross interference to the other spreading factor. As spreading factors 10-12 have low interference thresholds over the others, there is no need to lower their transmission power either as their transmissions are less likely to power-suppress other transmissions. However, spreading factors 8 and 9 have only about $-13 dBm$ isolation threshold against spreading factors 10, 11 and 12. Therefore, \emph{FREE} lowers their transmission power by $1 dBm$ so as to avoid cross spreading factor collisions with transmissions on the same channel. This allocation technique almost eliminates all cross collisions among transmissions, which improves the overall packet delivery ratio and energy consumption of the network significantly.

\begin{table}[!ht]
\centering
\caption{Transmission Power and Channel Allocation} \label{txandchanl}
\begin{tabular}{c|c|c}
\textbf{Channels} & \textbf{Spreading Factors} & \textbf{Transmission Powers [dBm]} \\
\hline
\hline
1 & 7 & 14 \\
2 & 9, 10, 11, 12 & 13, 14, 14, 14 \\
3 & 8,11,12 & 13, 14, 14 \\
\end{tabular}
\end{table}

\subsection{Packet Lengths, Guard Periods, and Slots per Frame}

\begin{algorithm}[!t]
\LinesNumbered
\SetKwInOut{Input}{input}
\SetKwInOut{Output}{output}
\Input{$P_{f,PER}\forall f\in\mathcal{F}$,$N_f\forall f\in\mathcal{F}$,$MaxSize_f\forall f\in\mathcal{F}$,$M_f\forall f\in\mathcal{F}$,$d$,$H$,$I$,$V$,$Skewrate$}
\Output {$L_{f}\forall f\in\mathcal{F}$,$G_{f}\forall f\in\mathcal{F}$,$S_{f}\forall f\in\mathcal{F}$}
\For{$f \in \mathcal{F}$}{
$\begin{aligned}
L_f=\argmin_{4<l<255}(1+R_f)\ceil{\frac{MaxSize_f}{l-H}}T_l I V \nonumber
\end{aligned}$
}
\For{$f \in \mathcal{F}$}{
$G_f = \ceil{Skewrate\big[\max(X_{f},\ceil{1/d}) \ceil{\frac{MaxSize_f}{L_{f}M_{f}}}+(M_{f}-1)\big](T_f)}$\;
$minS_{f}=\ceil{(\frac{T_f}{d})/(T_f+2G_f)}$\;
$S_f = N_f$\;
\If{$S_f < minS_f$}{
$S_f\,\,=minS_f$\;
}
}
\Return{$L_{f}\forall f\in\mathcal{F}$,$G_{f}\forall f\in\mathcal{F}$,$S_{f}\forall f\in\mathcal{F}$}\;
\caption{Slot lengths and Number of Slots}\label{slandsn}
\end{algorithm}

During the first stage, transmission parameters (spreading factor, transmission power, channel(s), and slot number) are assigned to each device separately. However, the frame structure and specifically, the number of slots per frame and the actual slot length, are not known by devices at that point because this information requires knowledge of the spreading factor distribution among all devices. At the end of the first stage, the gateway will have complete knowledge of all devices that participate in the coming data collection. We have designed Algorithm~\ref{slandsn} to determine the frame structure, which is broadcast periodically in the \textit{FSettings} messages as part of the second stage.

Here, the slot length depends on the packet transmission and guard times, and is equal to the transmission time plus 2 times the guard time. The packet lengths should be selected as explained in Section~\ref{pcktlen} to minimize energy consumption. All devices that use the same spreading factor need to have the same slot length to maintain synchronization. Therefore, the slot length is derived from the longest packet length, which, in turn, is computed based on the maximum buffered data size for a particular spreading factor (line 2). Once the packet length is known, it is used to calculate the guard period which should be large enough to accommodate the maximum clock skew accumulation estimate during the previous data collection round (line 5).

With the guard time and the slot length known, Algorithm~\ref{slandsn} calculates the minimum number of slots per frame ($minS_f$) that satisfies the duty cycle (line 8). If the actual number of slots per frame ($S_f$) is less than $minS_f$, it is set to $minS_f$. After receiving the frame structures in a \textit{FSettings} message, each device will know its slot length and the number of slots for its frame.

\section{Performance Evaluation} \label{evaluation}

To validate our work, we developed a simulation tool, called LoRaFREE, using Simpy and used the log-distance path loss model of LoRaSim \cite{bor2016lora}, which was obtained from measurements in a real environment. LoRaFREE is more comprehensive than LoRaSim as it considers a packet error model, the imperfect orthogonality of spreading factors, the fading impact, and the duty cycle limitation at both, the devices and the gateway. In addition to that, LoRaFREE supports bidirectional communication by adding the downlink capability and a re-transmission strategy in case of confirmable uplink transmissions. Furthermore, we extended the energy consumption profile from LoRaSim to consider the consumed energy at reception time. The aforementioned features are required for a proper evaluation of LoRa-based systems making LoRaFREE beneficial to the research community\footnote{\url{https://github.com/kqorany/FREE}}.

We are simulating and comparing results from the two \emph{FREE} scheduling scenarios (i.e. $\alpha=1$ and $\alpha=0$) with two other approaches, namely Legacy LoRaWAN and Delayed LoRaWAN, for both, unconfirmable and confirmable transmissions. In all scenarios, we consider an application that its rate of packet generation follows a Poisson distribution. The difference between the scenarios is in how and when those packets are sent to the gateway. Legacy LoRaWAN uses the standard LoRaWAN MAC, where devices follow the Class-A specification, i.e. Aloha-type MAC, two receive windows after each uplink, etc., and transmit immediately whenever data is generated by the application. In Delayed LoRaWAN, devices buffer the application data and transmit in bulk at known times. The devices, in this scenario, still follow the Class-A specification and do not perform any sort of synchronization before transmission. Only a time offset is used before the first transmission and then the rate of transmission is governed by the duty cycle of the channels. The purpose of the time offset here is to desynchronize devices in order to reduce systematic collisions. As the data is transmitted in bulk, devices send long packets to reduce the overhead of MAC headers as per Algorithm~\ref{slandsn} (lines 1-2). The spreading factor allocation in both the Legacy and the Delayed LoRaWAN scenarios is chosen to minimize the transmission time \cite{bor2016lora}. Finally, the devices in the two proposed \emph{FREE} scenarios follow the protocol presented in \S~\ref{sysmodel}, going through the joining and synchronization phase before starting the actual data transmission.

\begin{table}[!t]
  \centering
  \caption{Simulation Parameters} \label{simparameters}
\begin{tabular}{l|l}
	\textbf{Parameters} & \textbf{Value [Unit] + Comment} \\
    \hline    
    \hline
    Random Seeds & 10  \\
    Devices & 10 - 2000 Randomly scattered \\
    Bandwidth & 500 [KHz]\\
    Coding Rate & 4/5  \\
    Spreading Factor & 7-12 \\
    Concurrent Receptions & 8 \\
    Channels & 3 Uplink/Downlink with 1\% Duty Cycle \\
             & plus 1 Downlink with 10\% Duty Cycle \\
    Retransmissions & 8 Times Before Dropping \\
    LoRaWAN MAC Header & 7  [Bytes] \\
    \emph{FREE} MAC Header & 8 [Bytes]  Extra byte for Data Ordering \\
    LoRaWAN ACK & 0 [Bytes] Empty MAC header \\
    \emph{FREE} ACK & Depends on the number of slots \\ 
             & per frame \\
    Path Loss \cite{bor2016lora}& $\overline{L_{pl}}(d_0) = 127.41$[dB] \\
              &$d_0 = 40$[m], $\gamma=2.08$, $\sigma=2$ \\
    Application & 20 [Bytes] every exp(5 [mins]) \\
    Collection Periodicity & 1 - 48 [Hours] \\
    Transmission Offset & Uniform(0,600000) [ms] \\
    Battery Capacity & 1000 [mAh] \\
    Power (Transmission)  & 132 [mW] \\
    power (Reception)     & 48  [mW] \\
    Clock Skew Rate &  15  [$\mu$sec] every [sec]\\
    SNR Limits & Table~\ref{snrlimts} \\
    Receiver Sensitivities & Equation~\ref{sensitivity} \\
    NF & 6 [dB] \\
    Packet Error Model & Equations~\ref{ber} -~\ref{per} \\
    \end{tabular}
\end{table}

\begin{figure*}[!t]
  \centering
  \subfloat[Energy Consumption]{
    \includegraphics[width=0.7\columnwidth]{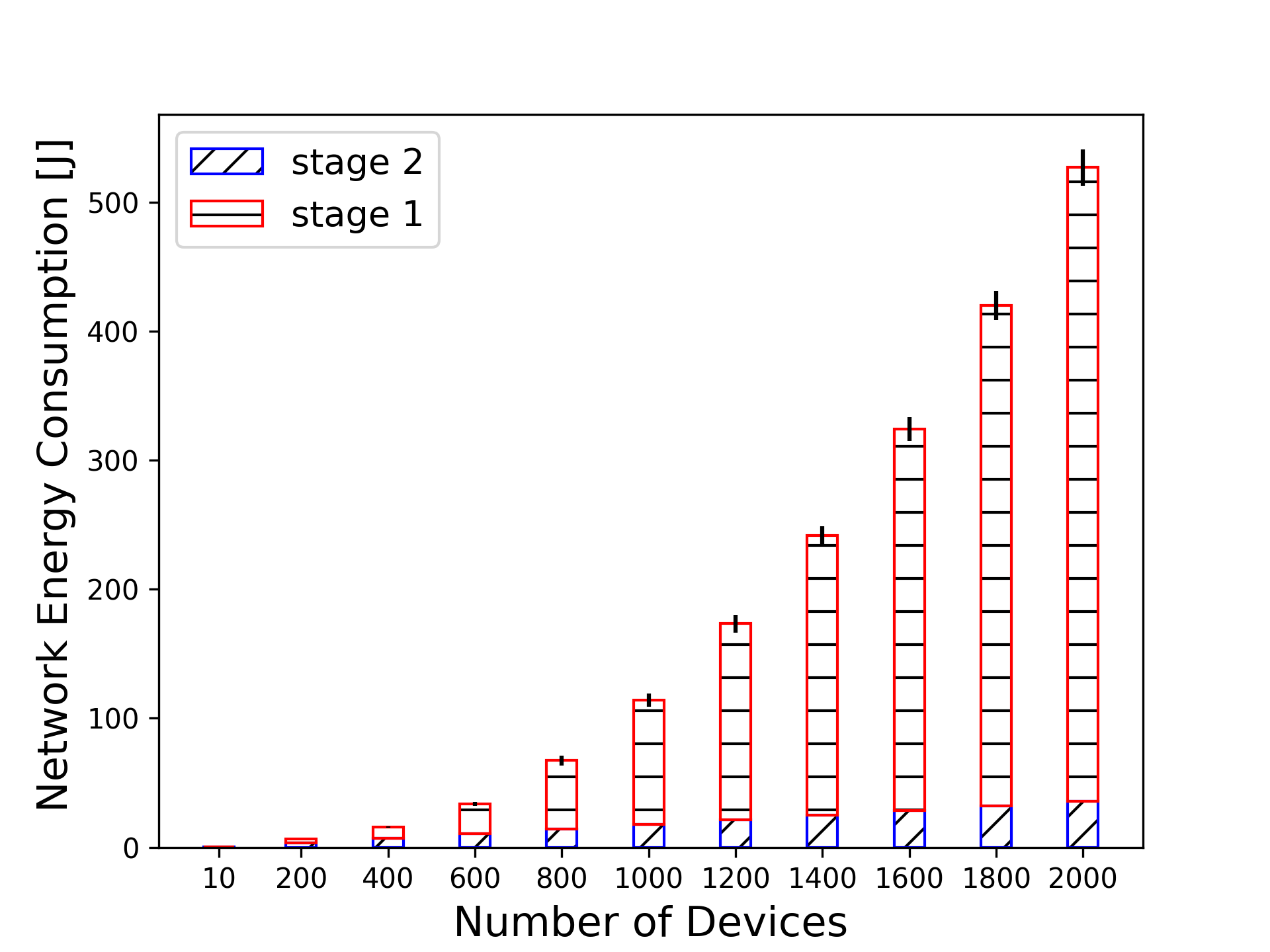}\label{s1and2energy}
  }
  \subfloat[Length in Time]{
    \includegraphics[width=0.7\columnwidth]{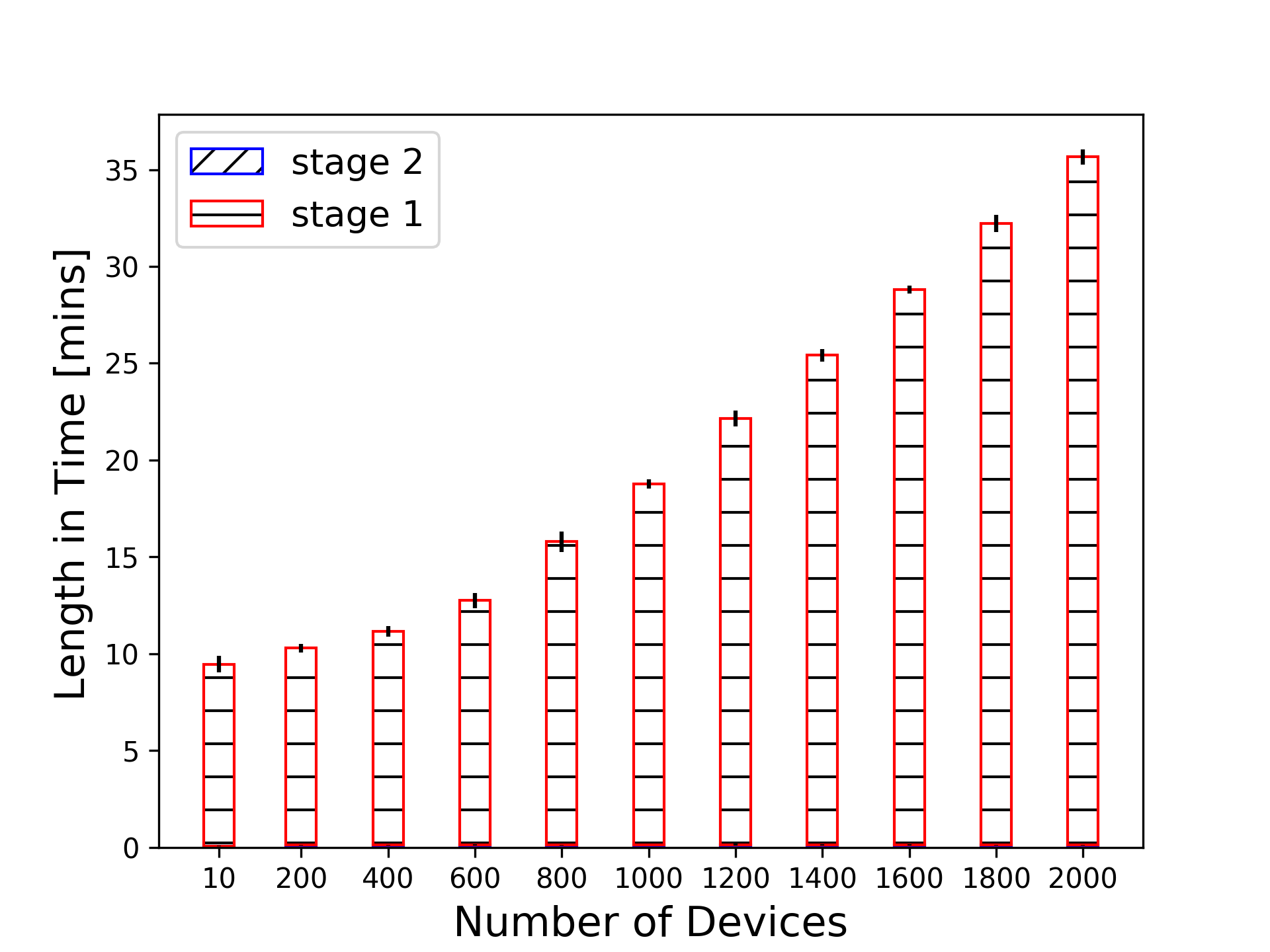}\label{s1and2period}
  }
  \subfloat[Statistics from Stage 1]{
    \includegraphics[width=0.7\columnwidth]{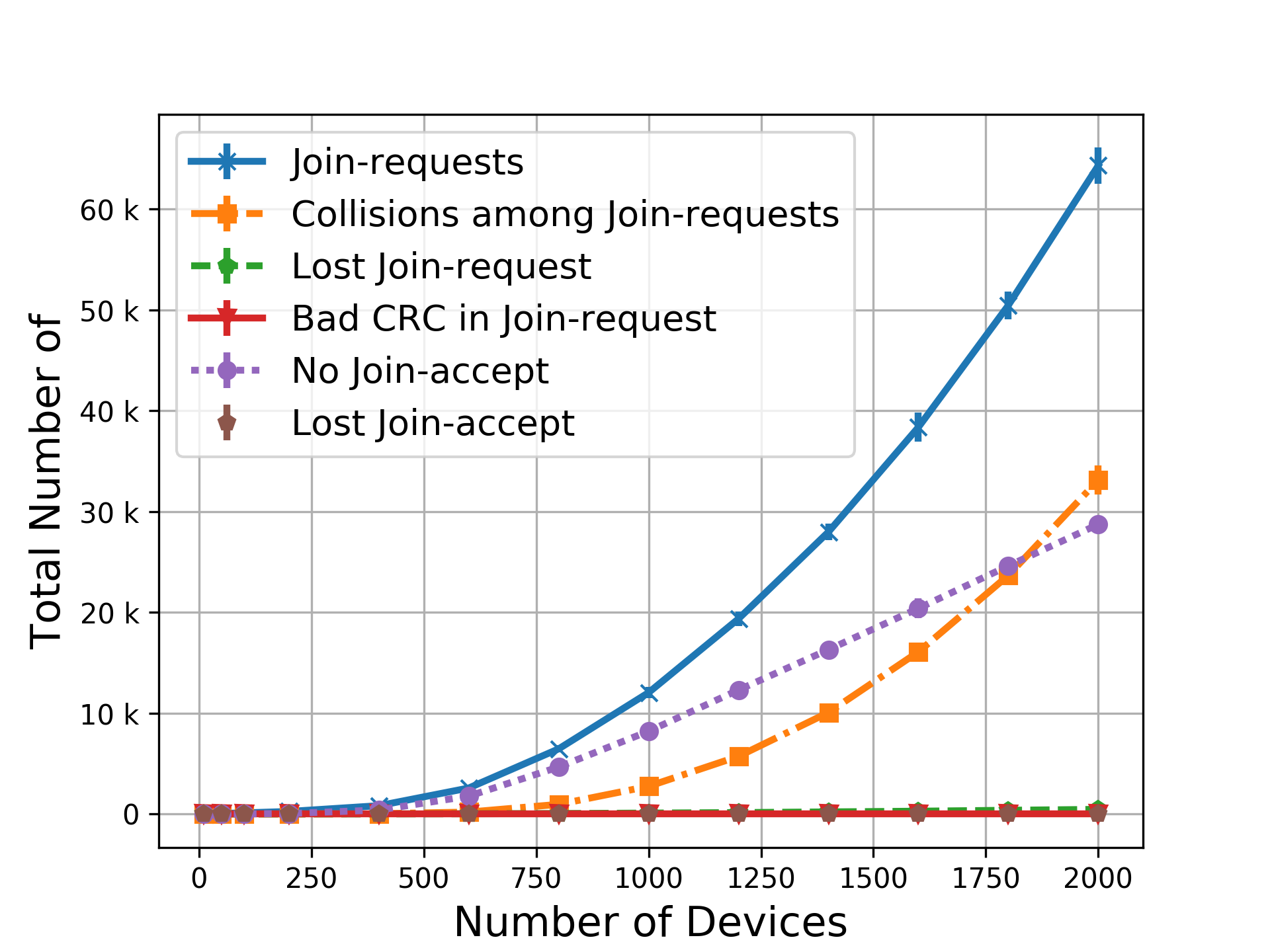}\label{s1stats}
  }
  \caption{Joining and Synchronization Phase Study}\label{s1and2study}
  \vspace{-0.35cm}
\end{figure*}

The scenarios are simulated with a fixed collection period, e.g. one day. In the simulation, all devices use a buffer of data (termed goal buffer) from which packets are extracted and sent to the gateway. The initial size of the goal buffer represents the total amount of application data generated during a collection period. The simulation terminates when all devices have transmitted all data in their goal buffers. For instance, if data is collected once every day and $20 B$ of application data are generated every $5 mins$, each device would start with $5760 B$ to be transmitted. Table~\ref{simparameters} summarizes all the simulation parameters we used in our evaluation. Each simulated study is executed 10 times using different random seeds and the mean across all results is presented along with the standard deviation.

We present the evaluation results for all scenarios in terms of:
\begin{itemize}
    \item Network energy and data collection time, where minimizing the two metrics helps the objective of this work.
    \item Estimated device lifetime, which can be deduced from the network energy consumption.
    \item Network data delivery ratio, which reflects the throughput of the network
\end{itemize}
For both \emph{FREE} scenarios, we show the spreading factor distribution, guard times, and the frame lengths in time. These results help to understand the performance of the two \emph{FREE} scenarios. Additionally, we introduce the air time efficiency metric to indicate the efficiency of the used schedule. This is computed as the ratio between the ideal and the actual data collection time\footnote{The ideal \emph{FREE} data collection would have each node transmitting only once without any guard periods or idle times, and with frames of the same length. Considering the use of 8 concurrent transmissions, the ideal data collection time length would then be the total time spent transmitting (over all the nodes) divided by 8.}. The metric reflects the sub-optimality introduced by the guard time as well as by the greedy online spreading factor allocation. Finally, the network statistics, i.e. transmissions, collisions, and lost packets, are presented to gain more insight into the results.

The results from the joining and synchronization phase are presented separately in \S~\ref{jandsps}. These results provide an insight into the overhead of this phase relative to the network size. Subsequently, performance results for the unconfirmable and confirmable data transmission studies are presented in \S~\ref{uas} and \S~\ref{cas}, respectively. The results of the two transmission types are obtained with a one-day data collection periodicity. Finally, the results obtained by varying the data collection periodicity to serve different data delay elasticity are presented in \S~\ref{dcps}.

\subsection{Joining and Synchronization Phase Study} \label{jandsps}

We investigate here the energy consumption (see Fig.~\ref{s1and2energy}) and the time required (see Fig.~\ref{s1and2period}) for devices to go through the joining and synchronization phase. The energy consumption and time required for the second stage increases \textit{linearly} with the network size. This is due to devices performing only receiving activities, which are negatively affected only by losing \textit{Fsettings} messages due to channel fading. In the first stage, energy consumption increases \textit{exponentially} and the time required for the stage increases \textit{supralinearly} with the the network size. The reason behind this is the scalability issue of the Aloha MAC and the duty cycle limitation of the gateway as shown in Fig.~\ref{s1stats}. For instance, in a network with $2000$ devices, the average number of transmitted \textit{join-requests} is $33$ per device. That is because roughly $17$ of these requests on average collide with other \textit{join-requests} and almost $14$ of these requests on average are received by the gateway. However, the gateway cannot send back \textit{join-accept} messages in both receive windows due to its limited duty cycle (see \textit{No join-accept} in Fig.~\ref{s1stats}). In addition to that, a very small percentage of \textit{join-requests} and \textit{join-accepts} are lost due to channel fading.

\subsection{Unconfirmable Traffic Type Study} \label{uas}

\begin{figure*}[!t]
  \centering
  \subfloat[Network Energy Consumption]{
    \includegraphics[width=0.7\columnwidth]{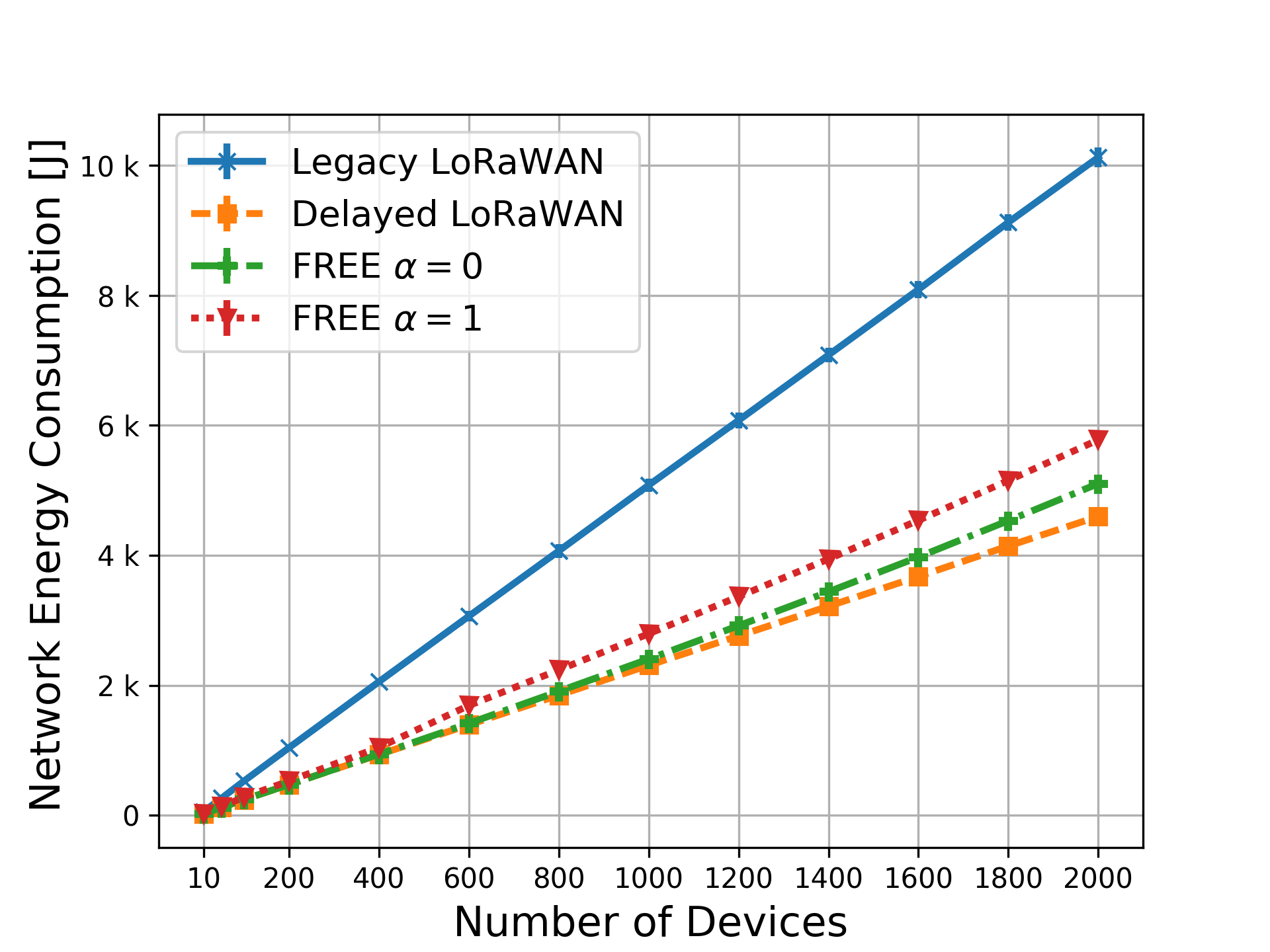}\label{Uenergy}
  }
  \subfloat[Device Lifetime]{
    \includegraphics[width=0.7\columnwidth]{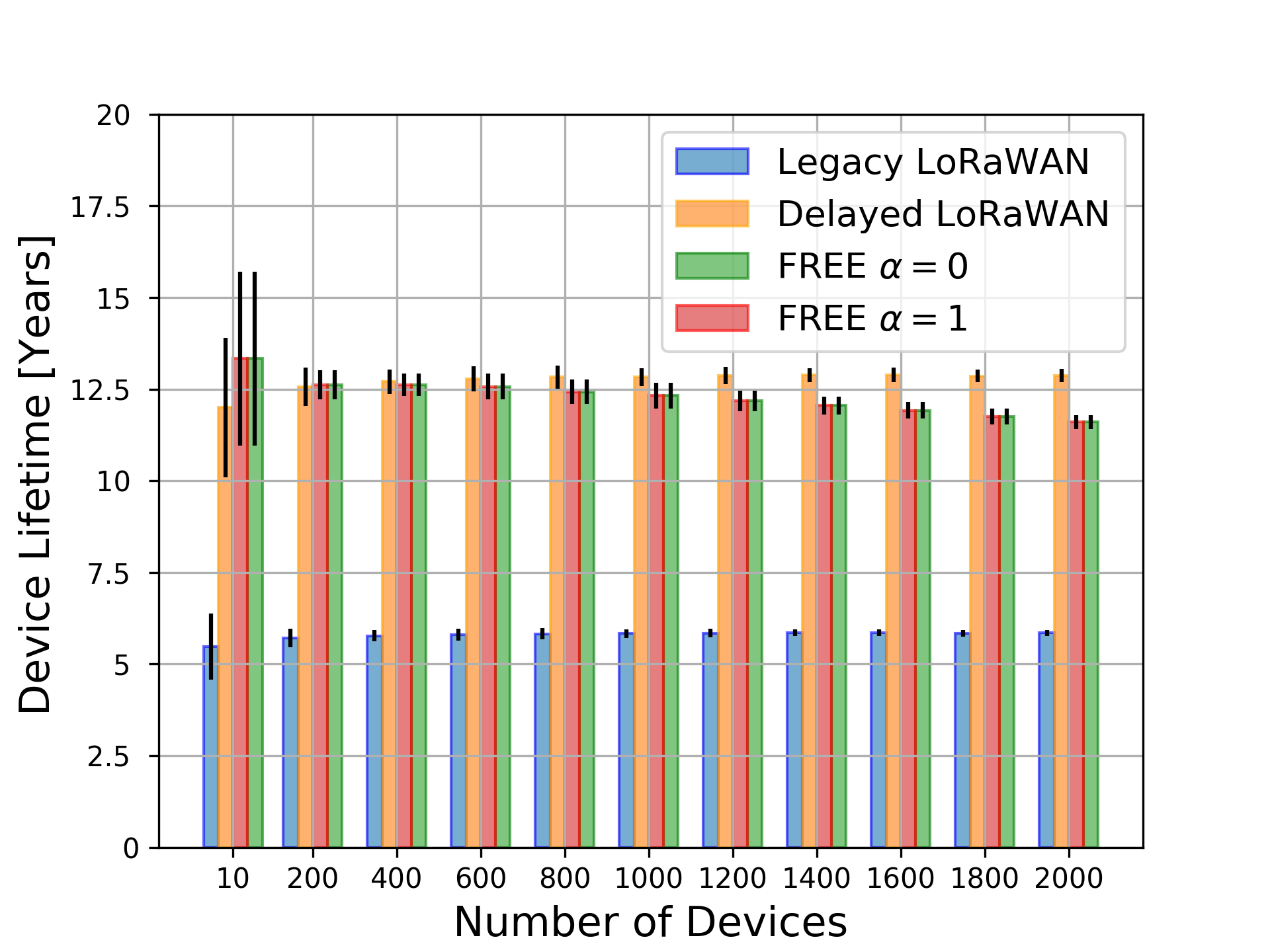}\label{Ulife}
  }  
  \subfloat[Uplink Transmissions]{
    \includegraphics[width=0.7\columnwidth]{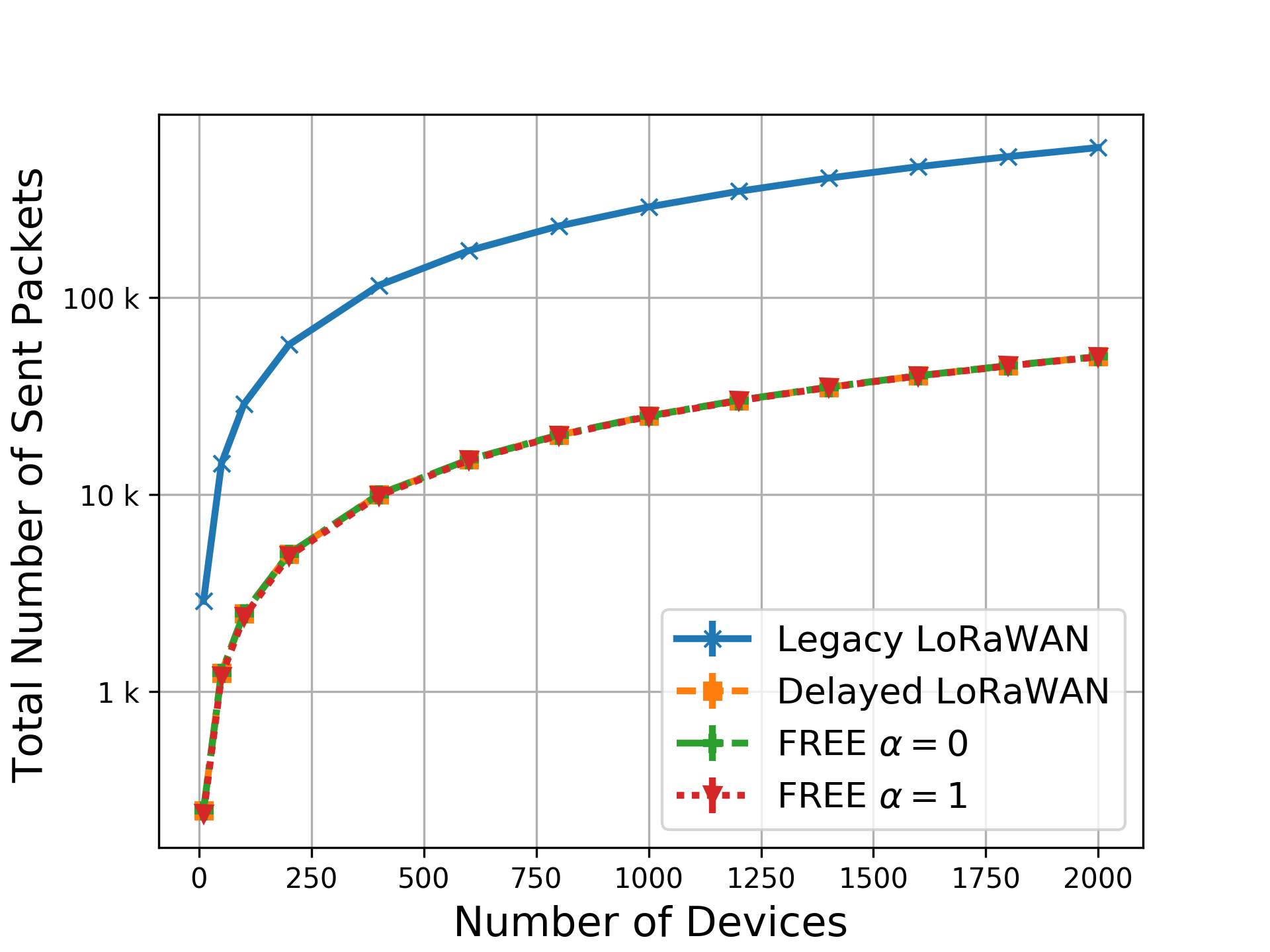}\label{Utrans}
  }
  \vspace{-.1ex}  
  \subfloat[Collection Time]{
    \includegraphics[width=0.7\columnwidth]{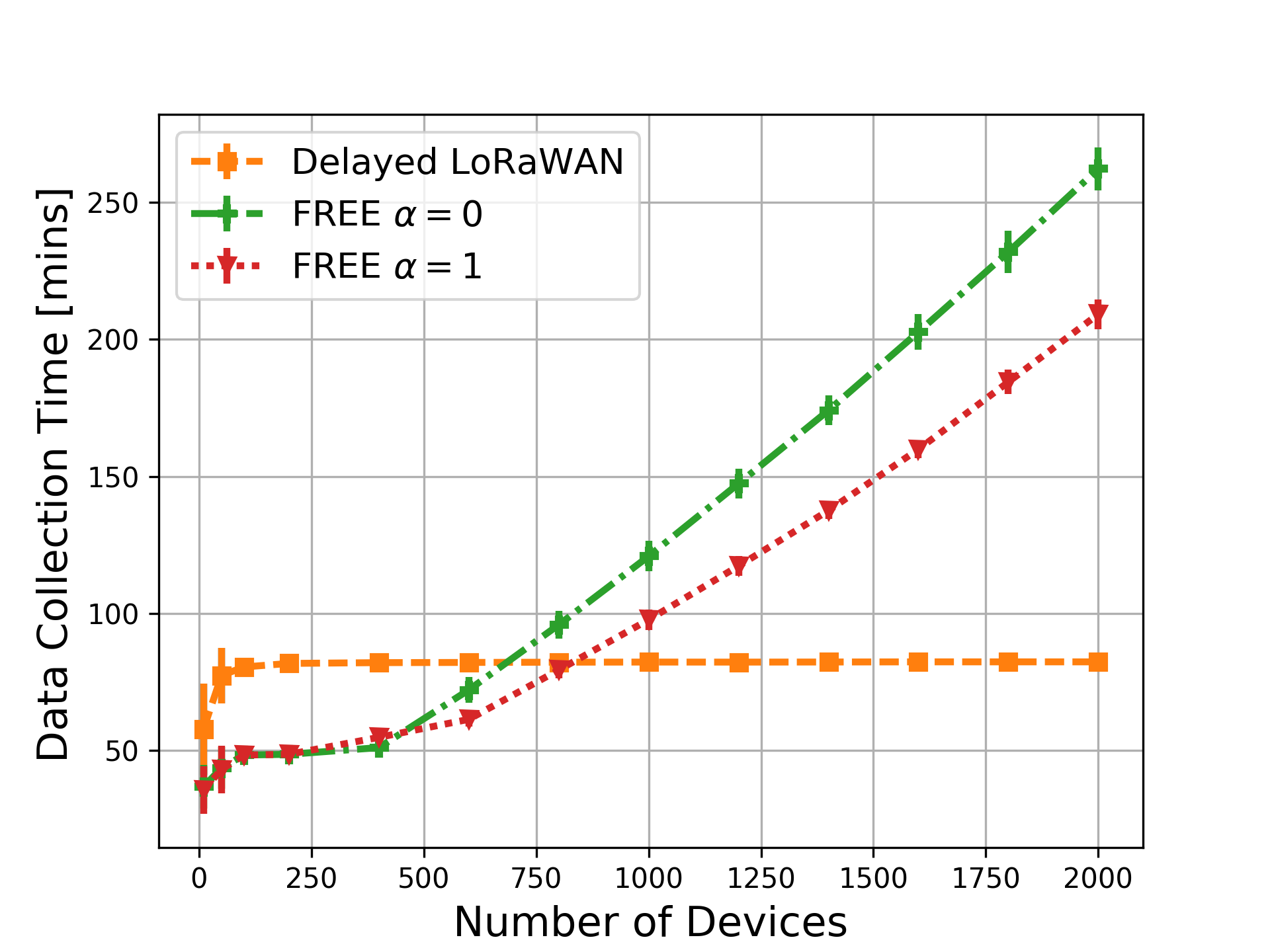}\label{Utime}
  }
  \subfloat[Spreading Factors]{
    \includegraphics[width=0.7\columnwidth]{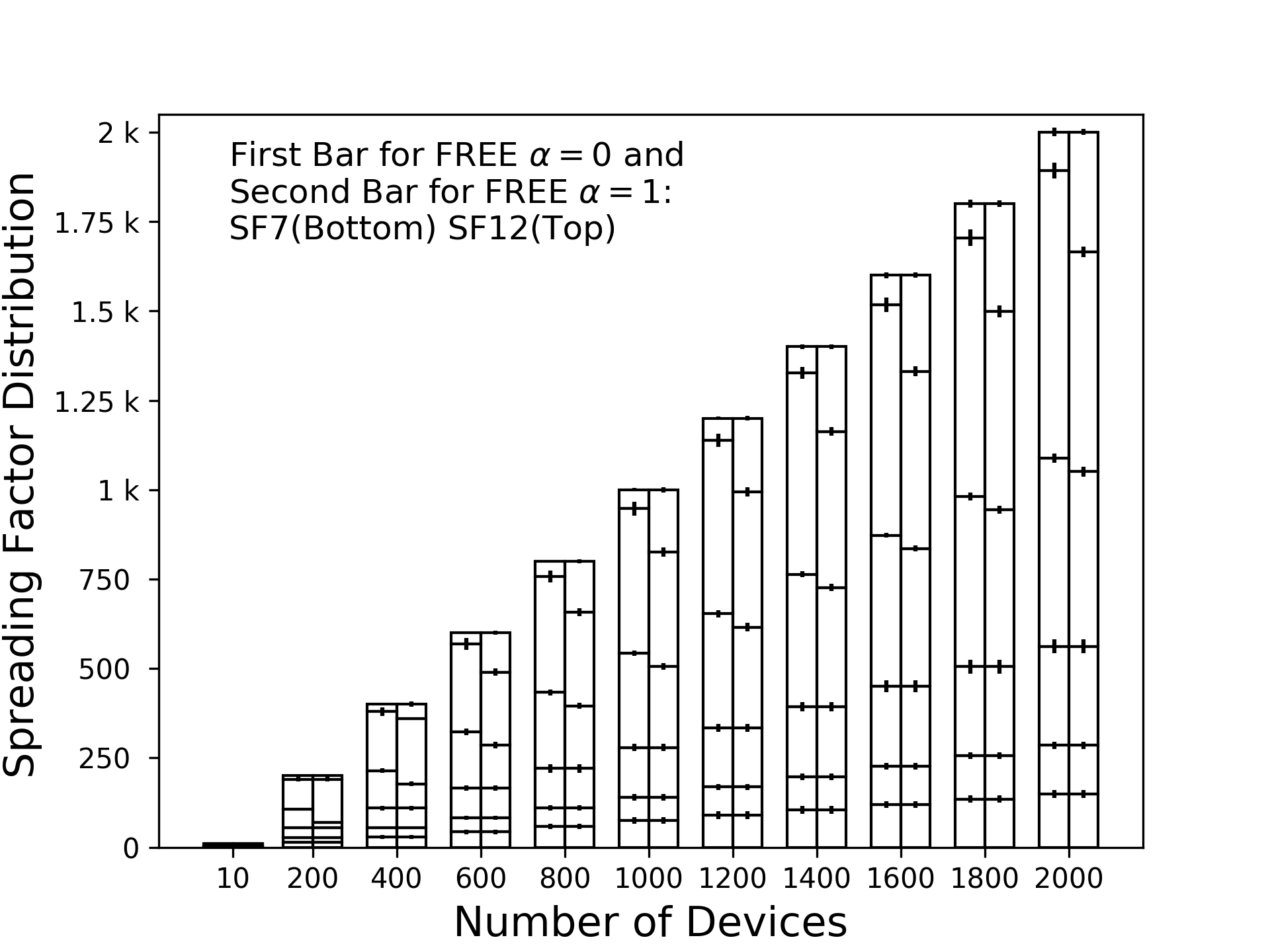}\label{Usf}
  }
  \subfloat[Guard Times per one Slot]{
    \includegraphics[width=0.7\columnwidth]{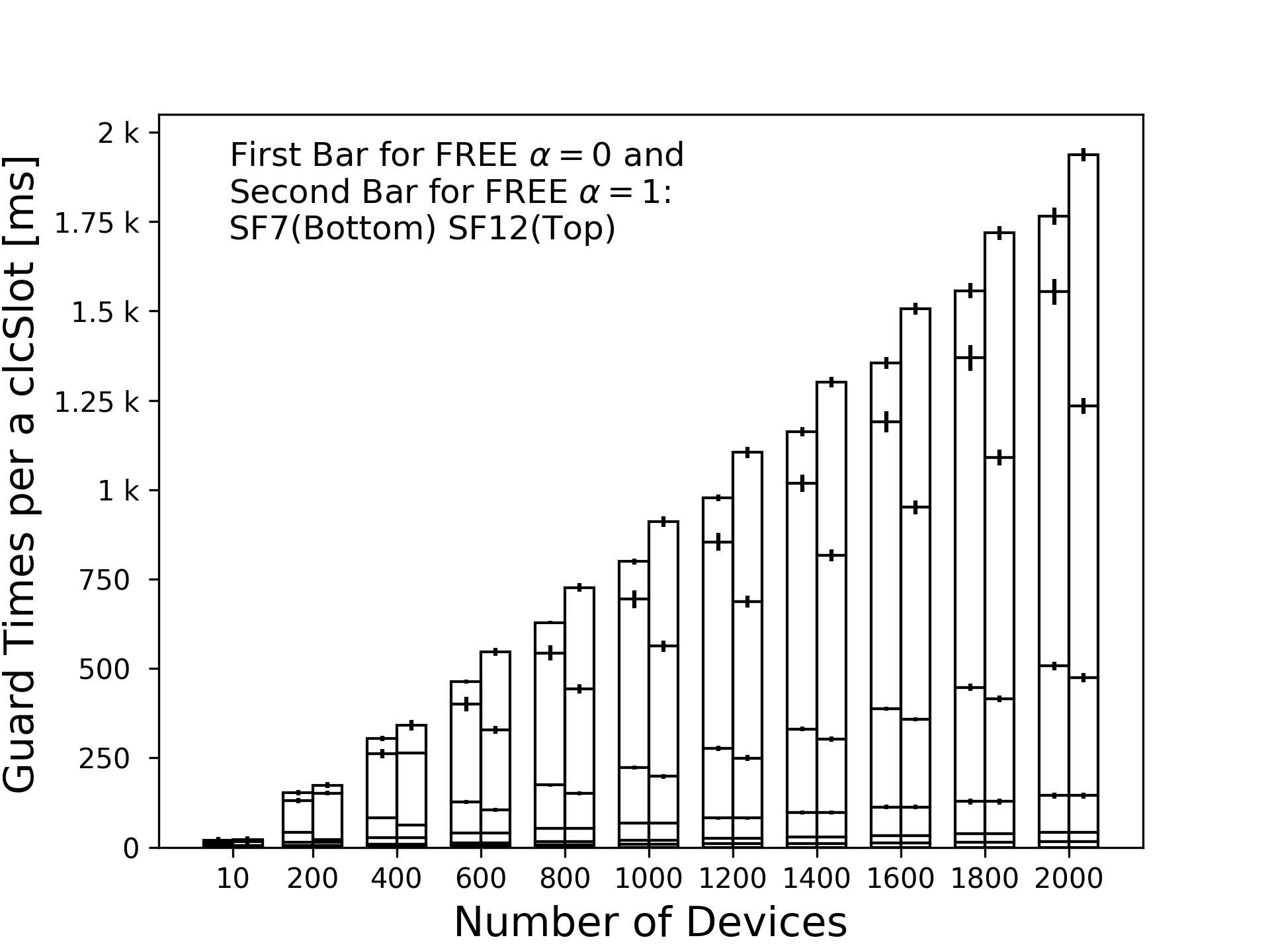}\label{Uguard}
  }
  \vspace{-.1ex}  
  \subfloat[Data Delivery Ratio]{
    \includegraphics[width=0.7\columnwidth]{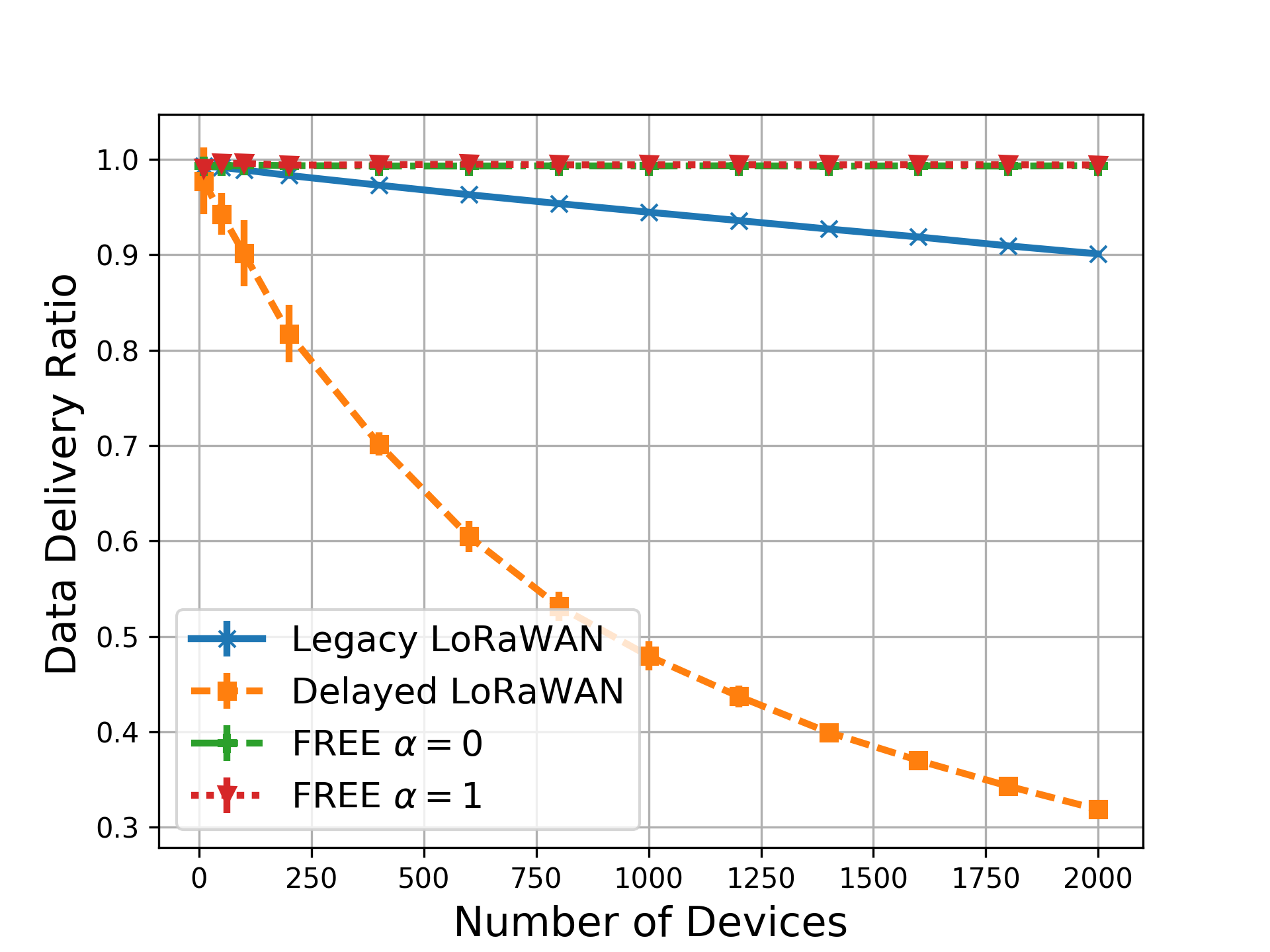}\label{Uder}
  }
  \subfloat[Uplink Collisions]{
    \includegraphics[width=0.7\columnwidth]{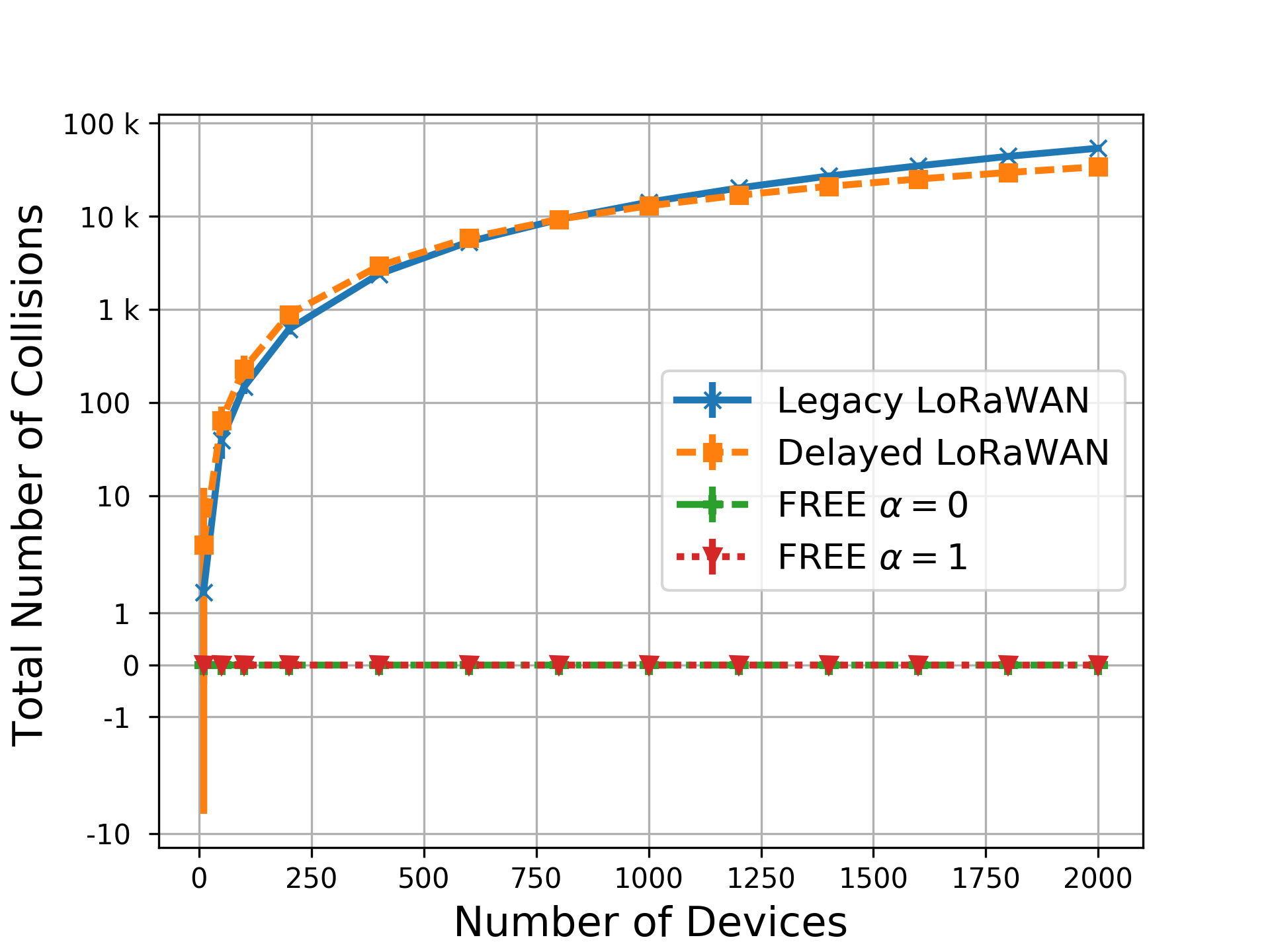}\label{Ucoll}
  }  
  \subfloat[Uplink Lost]{
    \includegraphics[width=0.7\columnwidth]{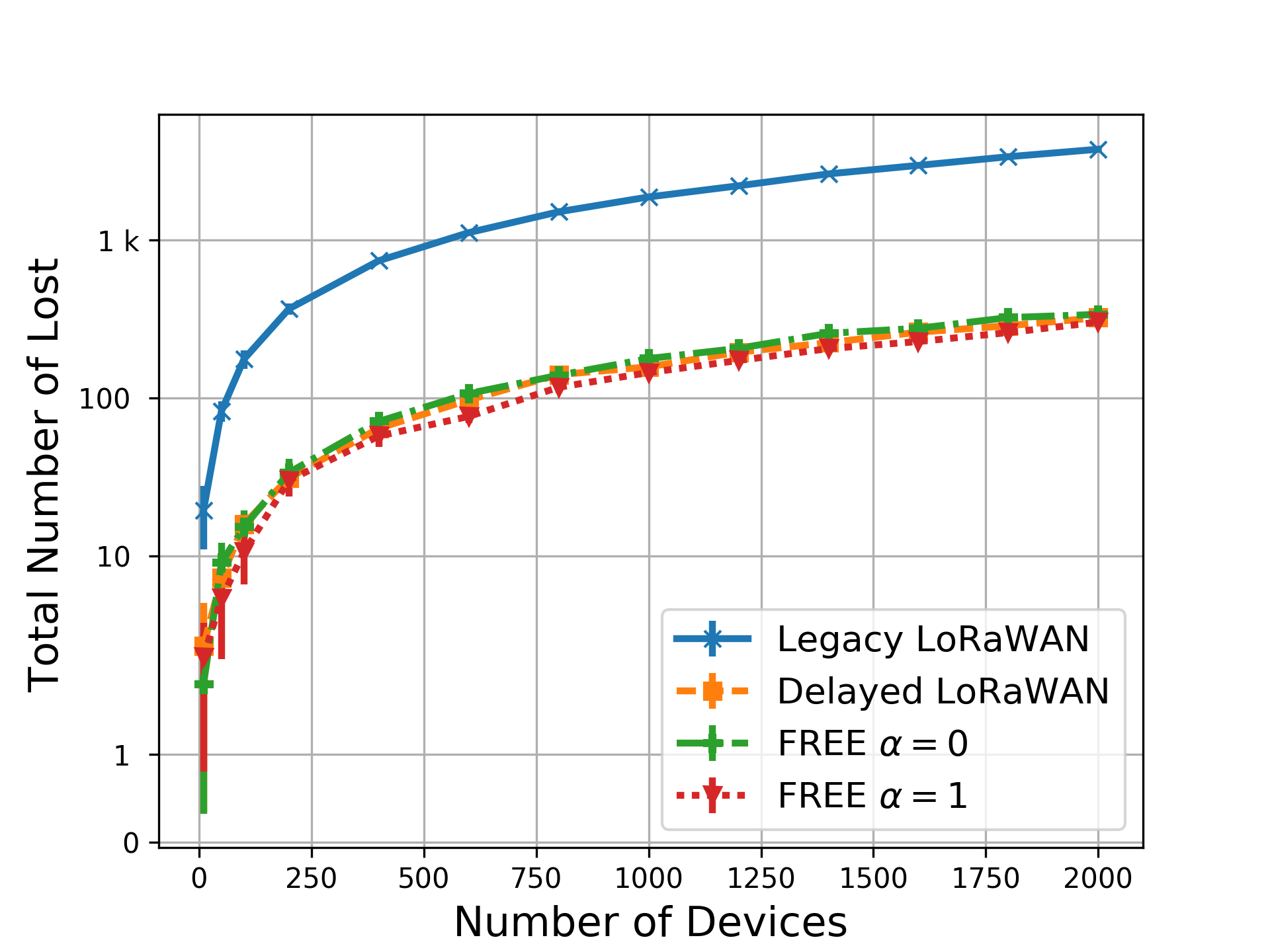}\label{Ulost}
  }
  \vspace{-.1ex}  
  \subfloat[Frame Lengths]{
    \includegraphics[width=0.7\columnwidth]{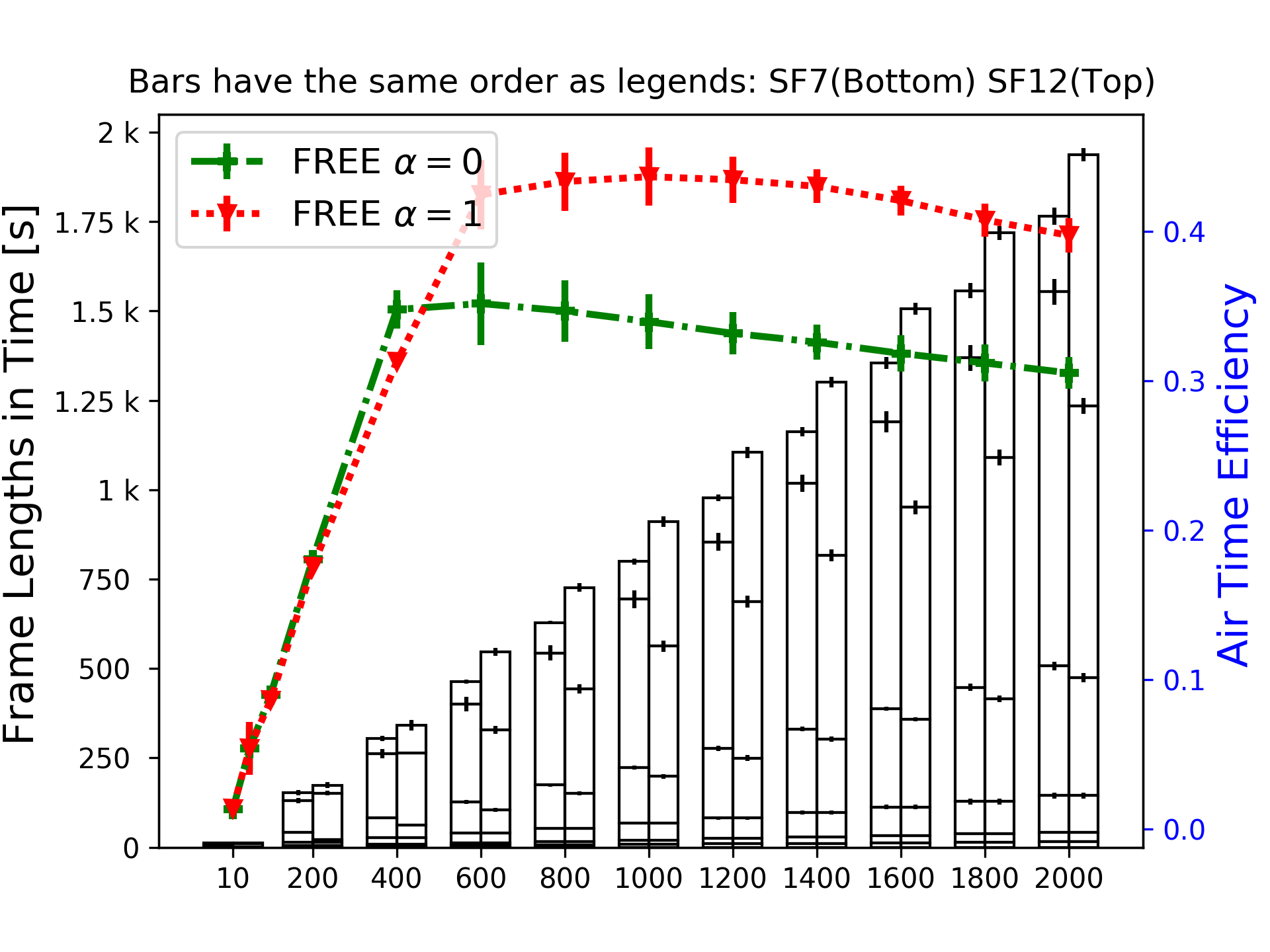}\label{Uframelength}
  }
  \caption{Unconfirmable Traffic Type Study}\label{unconfirmablestudy}
  \vspace{-0.35cm}
\end{figure*}

In the following, we study the unconfirmable traffic, where uplink transmissions are not acknowledged. The results of this study are presented in Fig.~\ref{unconfirmablestudy} for an application that generates 20 bytes from a Poisson distribution of 5 minutes rate and with a one-day data collection period.

Fig.~\ref{Uenergy} shows the network energy consumption for one day and Fig.~\ref{Ulife} shows an estimation of the device's lifetime. The lifetime estimation is calculated assuming a battery capacity of $1000 mAh$. Overall, both \emph{FREE} schemes (including the joining and the synchronization phase overhead) consume less energy than Legacy LoRaWAN, but show a higher energy consumption than Delayed LoRaWAN. This is reflected in the device lifetime, where in case of Legacy LoRaWAN, devices survive about $6$ years compared to $10$ plus years using the other approaches (see Fig~\ref{Ulife}).

The network size has a \textit{linearly} increasing impact on the network energy consumption and no impact on the devices' lifetime except for the two \emph{FREE} schemes. This is because of the unconfirmable transmissions, where collisions and lost packets do not affect the results. However, the differences among the schemes are in the overhead due to MAC headers and the joining and synchronization phase. In Legacy LoRaWAN, devices do not buffer data but transmit it right away. Therefore, the number of the overall transmitted packets is large compared to the other schemes (see Fig.~\ref{Utrans}) and, thus, the impact of MAC header overhead. For instance, a network with $1000$ devices transmits roughly $288 K$ packets in case of Legacy LoRaWAN, i.e. $2 MB$ of MAC headers, whereas about only $25 k$ packets are transmitted in the other systems, i.e., $0.2 MB$ of MAC headers. This is the reason why the Legacy LoRaWAN scheme consumes a lot more energy compared to the other schemes.

The \emph{FREE} schemes transmit overall the same number of packets as Delayed LoRaWAN. However, because of the overhead of the joining and synchronization phase, Delayed LoRaWAN consumes less energy than the \emph{FREE} schemes. The difference in energy consumption among the two schemes increases with an increase in network size due to the scalability issue of the joining and synchronization phase (Section.~\ref{jandsps}). This is also the reason behind the slight degradation in the device's lifetime for the \emph{FREE} schemes.

\emph{FREE} with $\alpha=0$ minimizes the energy consumption as shown in Fig.~\ref{Uenergy} at the expense of the overall data collection time as shown in Fig.~\ref{Utime} and vice versa for $\alpha=1$. This is due to the different spreading factor distributions that each scheme applies in order to optimize the corresponding objective function as presented in Algorithm~\ref{sfah}. As shown in Fig.~\ref{Usf}, when $\alpha=1$ some devices with low spreading factors have been shifted to higher spreading factors to balance the frame lengths and achieve a lower collection time (see Fig.~\ref{Utime}). However, this leads to an increased use of higher spreading factors and, thus, a higher energy consumption overall (see Fig.~\ref{Uenergy}).

\begin{figure*}[!t]
  \centering
  \subfloat[Network Energy Consumption]{
    \includegraphics[width=0.7\columnwidth]{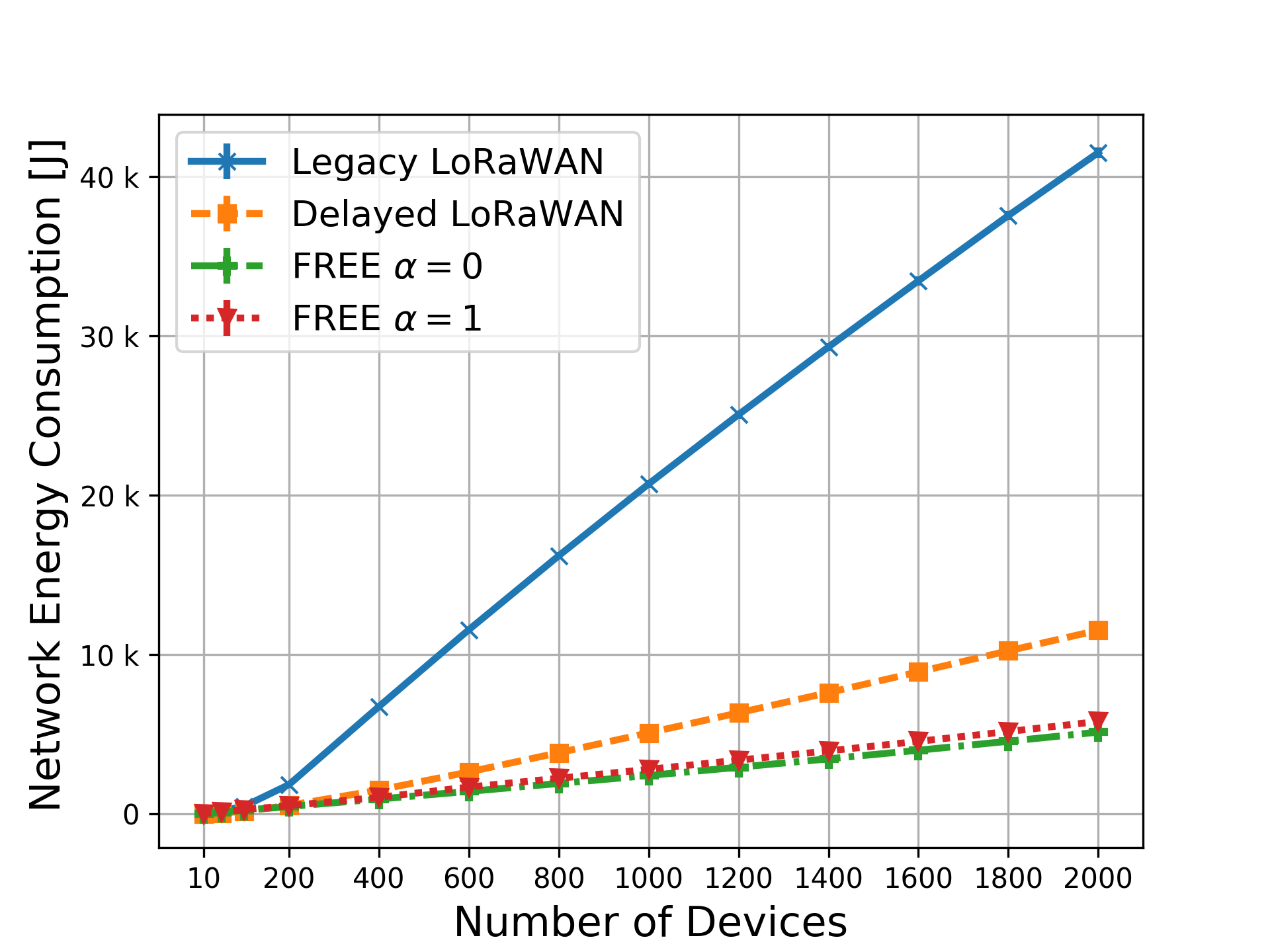}\label{Cenergy}
  }
  \subfloat[Device Lifetime]{
    \includegraphics[width=0.7\columnwidth]{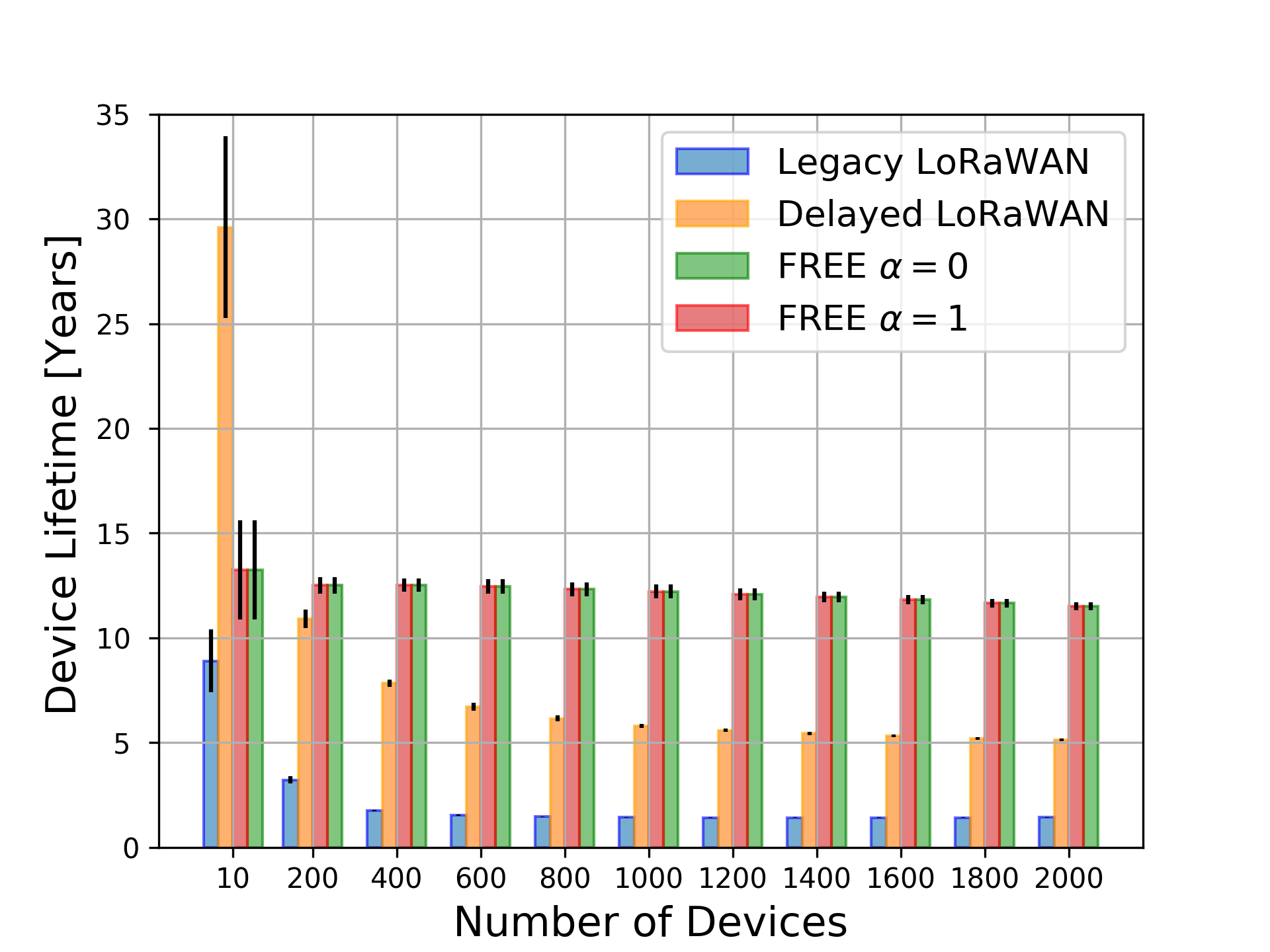}\label{Clife}
  }  
  \subfloat[Uplink Transmissions]{
    \includegraphics[width=0.7\columnwidth]{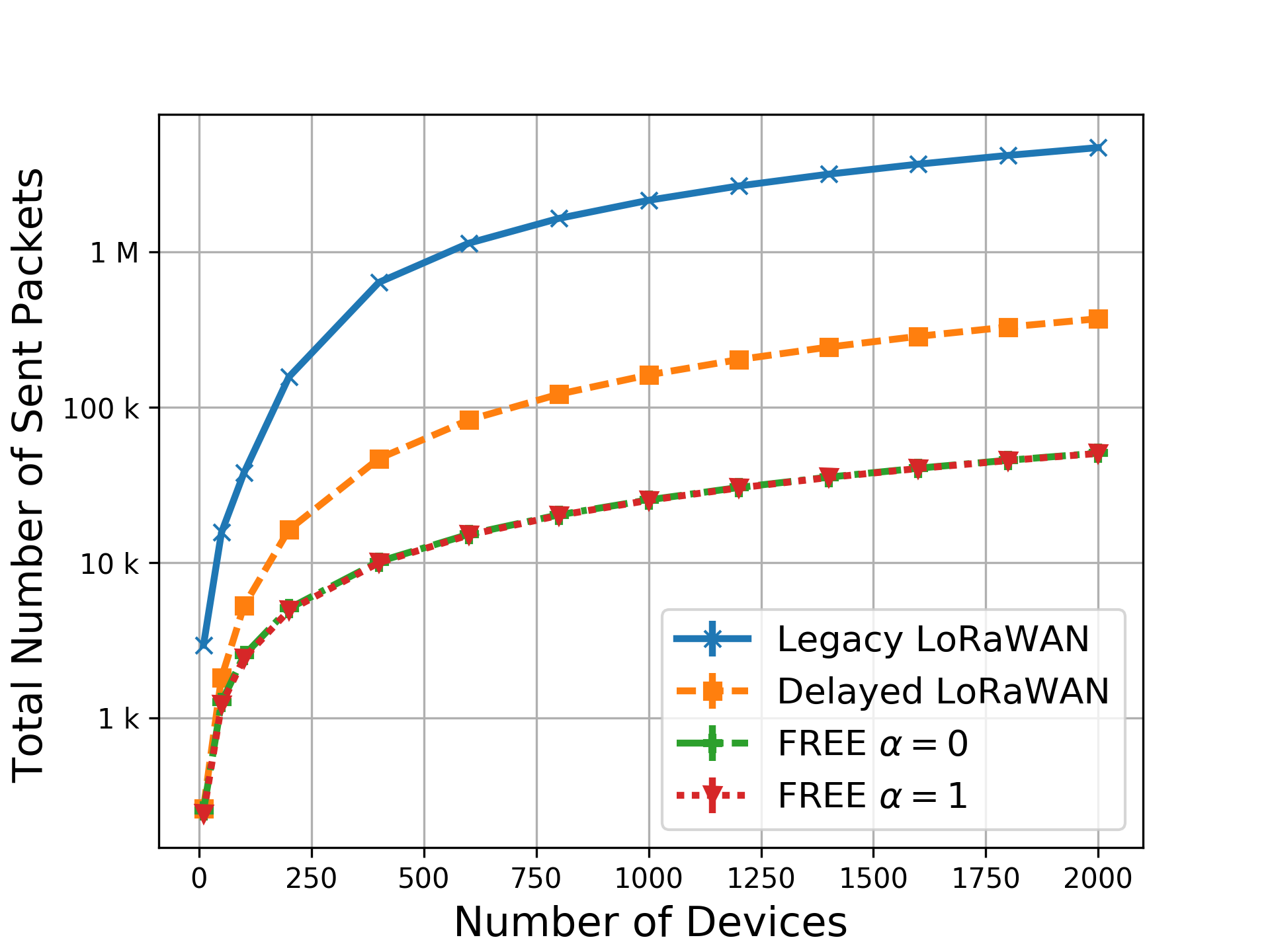}\label{Ctrans}
  }
  \vspace{-.1ex}  
  \subfloat[Collection Time]{
    \includegraphics[width=0.7\columnwidth]{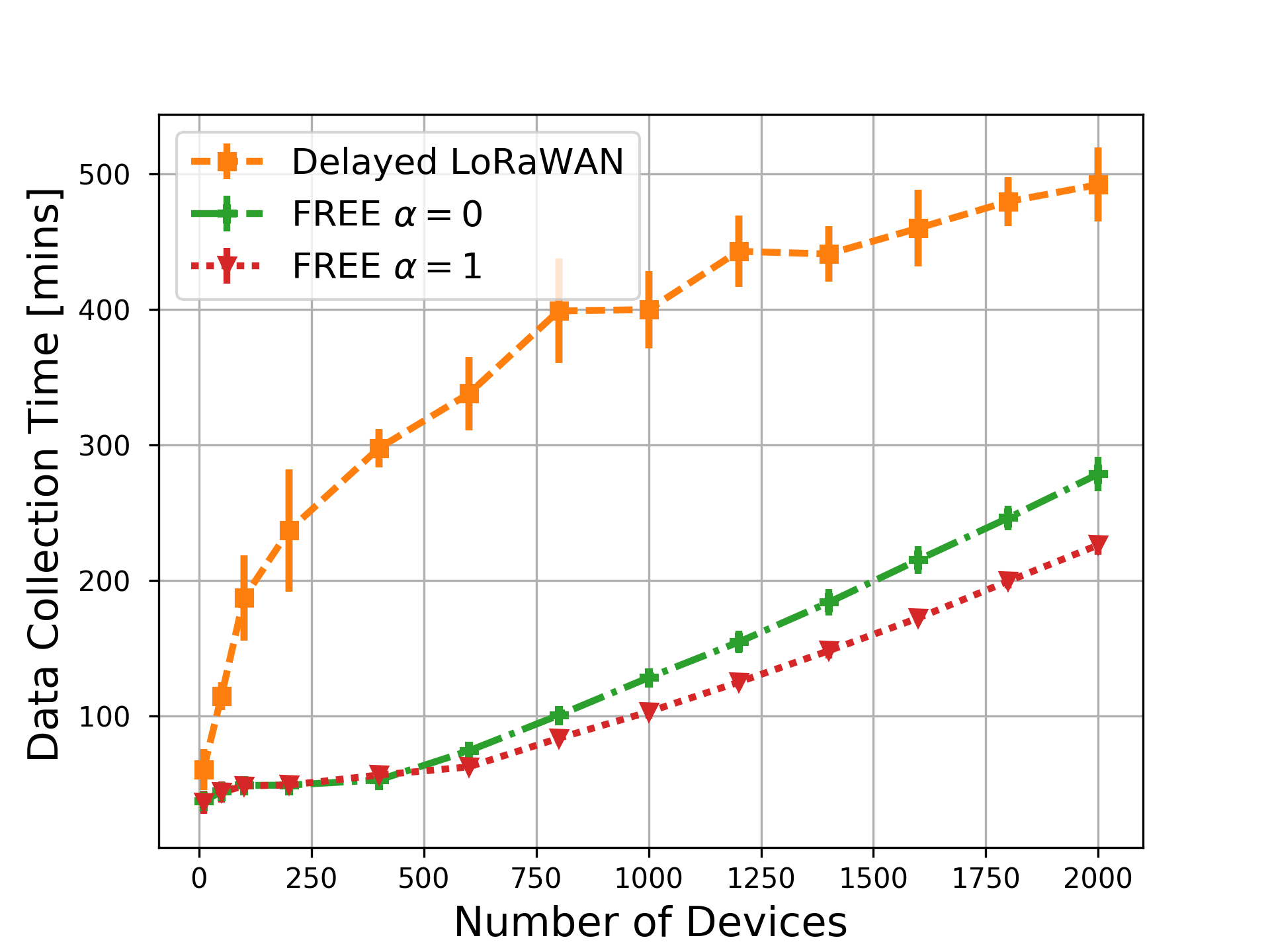}\label{Ctime}
  }
  \subfloat[Data Delivery Ratio]{
    \includegraphics[width=0.7\columnwidth]{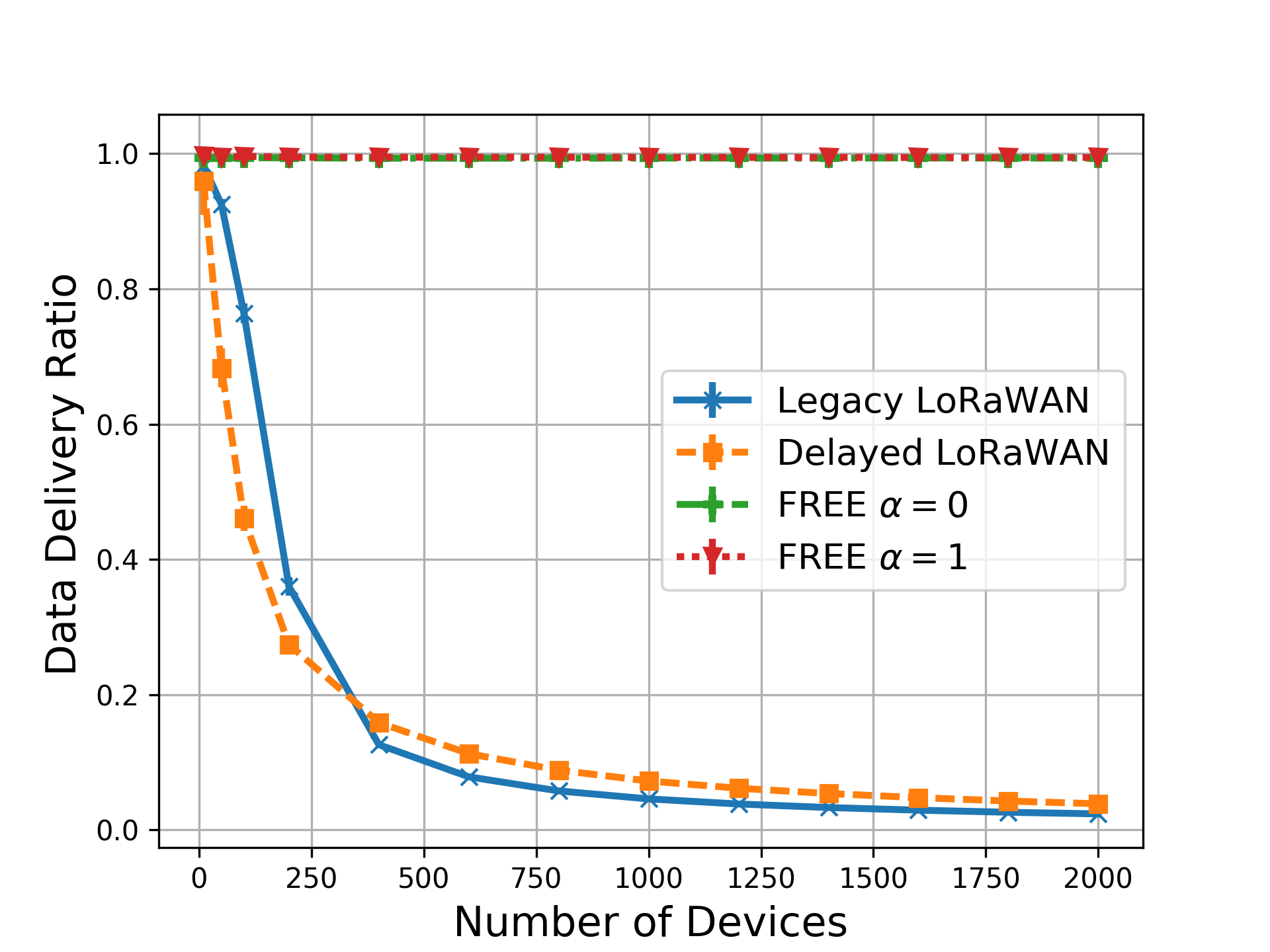}\label{Cder}
  }
  \subfloat[Uplink Collisions]{
    \includegraphics[width=0.7\columnwidth]{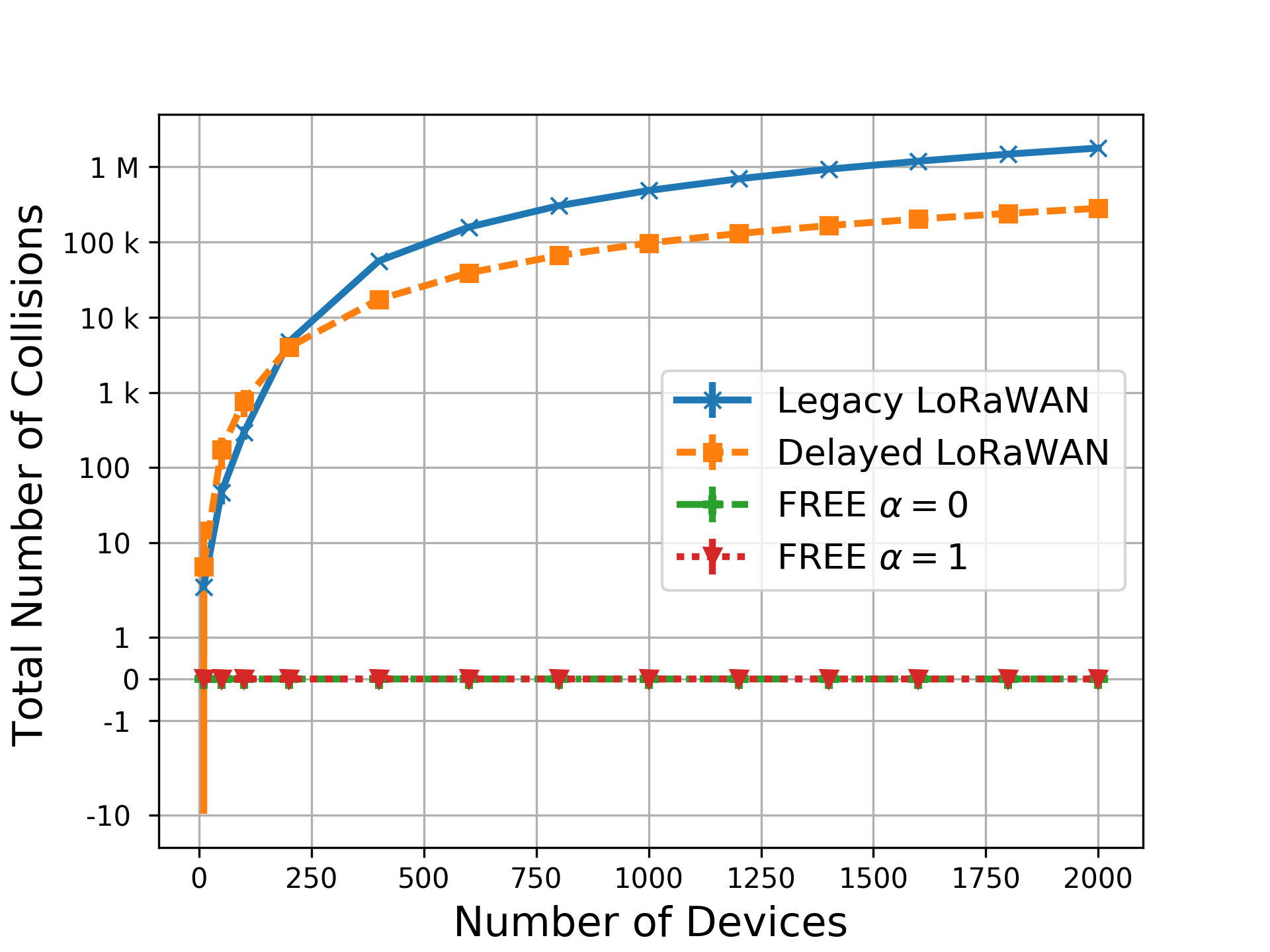}\label{Ccoll}
  }
  \vspace{-.1ex}
  \subfloat[Uplink Lost]{
    \includegraphics[width=0.7\columnwidth]{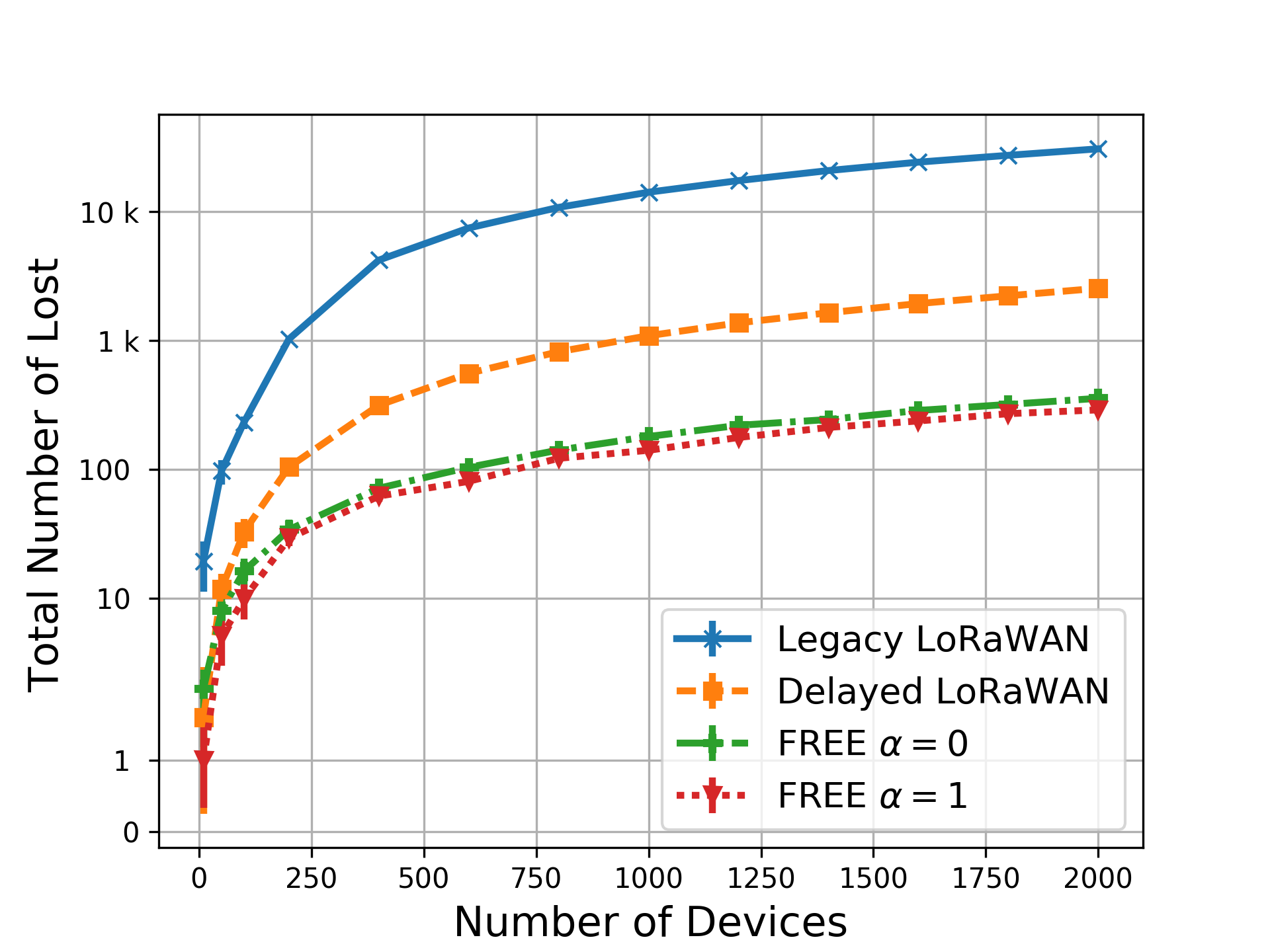}\label{Clost}
  }
  \subfloat[No ACK]{
    \includegraphics[width=0.7\columnwidth]{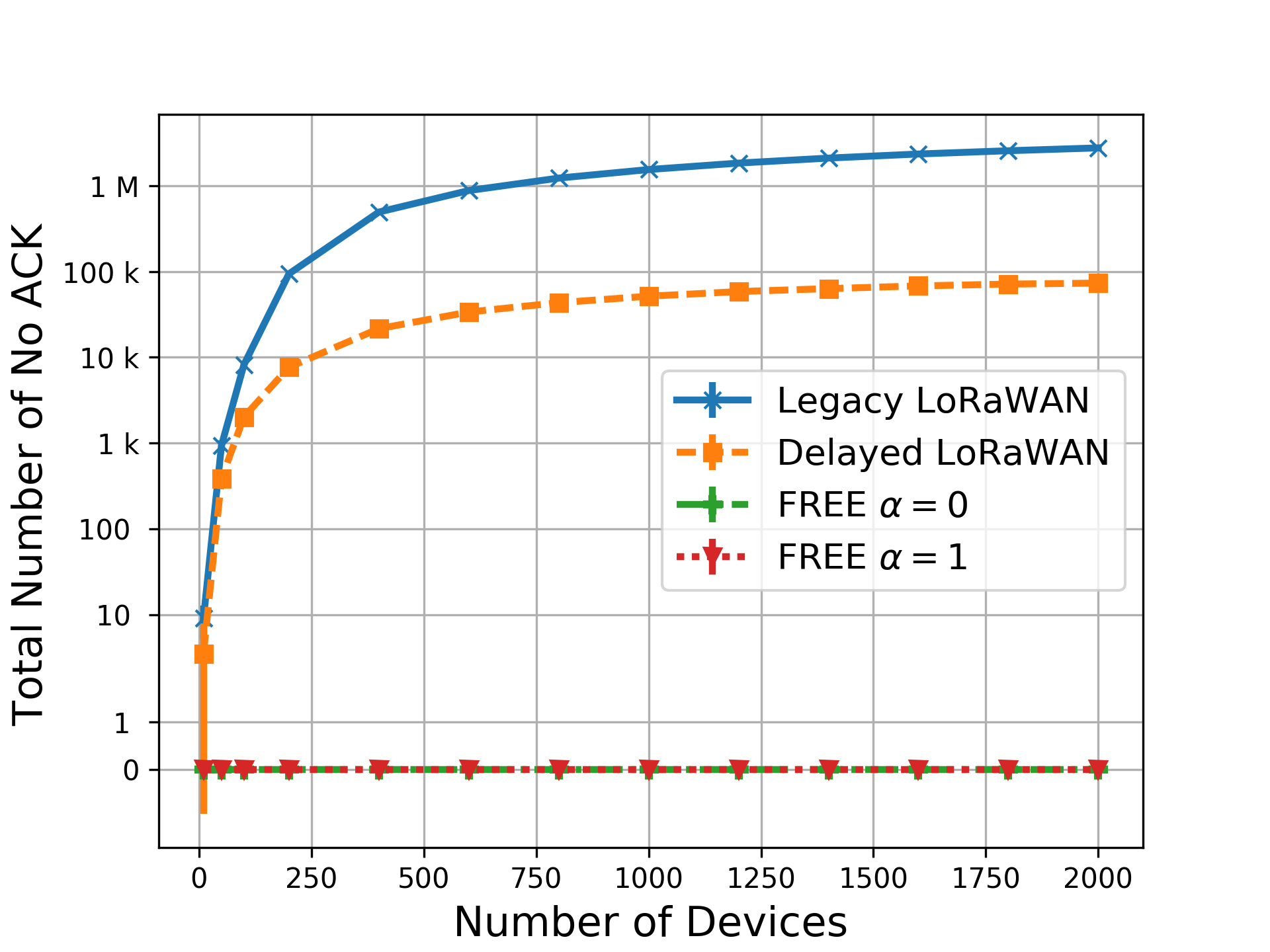}\label{Cnoack}
  }
  \subfloat[ACK Lost]{
    \includegraphics[width=0.7\columnwidth]{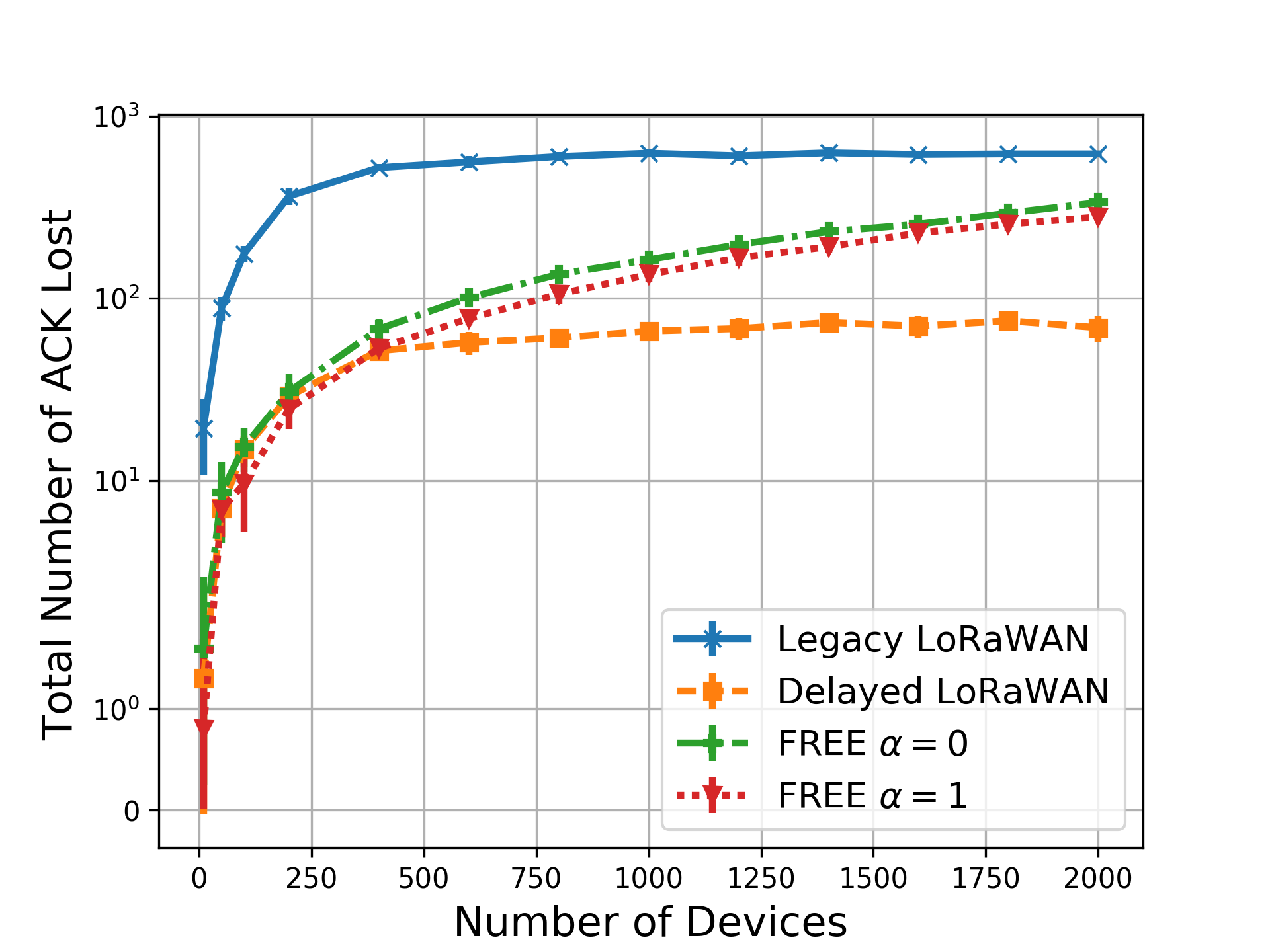}\label{Cacklost}
  }
  \caption{Confirmable Traffic Type Study}\label{confirmablestudy}
  \vspace{-0.35cm}
\end{figure*}

The lowest collection time is obtained with Delayed LoRaWAN\footnote{Collection time is not a relevant metric for Legacy LoRaWAN so it is not included in the comparison.}  (see Fig.~\ref{Utime}). There are two reasons for these results. The first reason is the time overhead of the joining and the synchronization phase and the second is due to the guard period used in each slot in \emph{FREE}. However, for small network sizes of less than $500$ devices, \emph{FREE} achieves less data collection time than Delayed LoRaWAN. This is because devices in Delayed LoRaWAN perform a transmission offset before the first transmission in order to alleviate collisions. Furthermore, for these network sizes, the guard periods are still small compared to the actual times used in transmitting in \emph{FREE} (see Fig.~\ref{Uguard}).

The low energy use and collection time achieved with Delayed LoRaWAN come at the expense of the Data Delivery Ratio (DDR) (see Fig.~\ref{Uder}). DDR represents the ratio of correctly received data to the initial buffer sizes of all devices. In the case of Delayed LoRaWAN, the DDR dramatically degrades by increasing the network size due to collisions (see Fig.~\ref{Ucoll}) and channel fading (see Fig.~\ref{Ulost}). Although Legacy LoRaWAN experiences a higher number of collisions and lost packets than Delayed LoRaWAN, it still achieves a higher DDR. That is because a collision or a loss is more costly in the case of Delayed LoRaWAN due to the large packet lengths. On the contrary, the effectiveness of synchronized communications in preventing most of the collisions in the \emph{FREE} scenarios results in a DDR of almost 1 regardless of the network size.

Fig~\ref{Uframelength} shows the frame lengths in seconds for all spreading factors in both \emph{FREE} schemes in addition to their air time efficiency. The frame lengths are directly proportional to the spreading factor distribution (Fig.~\ref{Usf}), the transmission times, and the guard periods (Fig.~\ref{Uguard}). As shown, the accumulated frame lengths increase for both schemes with increasing network size. In addition to that, the frame lengths associated with high spreading factors are mostly longer than the frame lengths of low spreading factors. That is because high spreading factors require more time to transmit packets of the same length than lower spreading factors. Consequently, high spreading factors require longer guard periods to accommodate clock skew as the time of their data collection rounds is longer than is the case for lower spreading factors (see Fig.~\ref{Uguard}).

The frame lengths are more balanced when $\alpha=1$ compared to the case of $\alpha=0$ and, thus, the air time efficiency is higher. It should be noted that the data collection rounds for spreading factors 11 and 12 are running on two channels at the same time. Therefore, roughly half the frame lengths of these spreading factors should be considered for a fair comparison with the other spreading factors. The balanced frame lengths result in a schedule that utilizes the concurrent transmissions as much as possible and, thus, minimizes the overall data collection time. This is the reason that the air time efficiency is higher when $\alpha=1$ than when $\alpha=0$. For small network sizes, the air time efficiency of both schemes is low because of the minimum frame length due to the duty cycle. As a result, even if only one device is assigned to a particular spreading factor, the corresponding frame length must be long enough to obey the duty cycle, leading to a lot of non-utilized time in the corresponding frame.

\subsection{Confirmable Traffic Type Study} \label{cas}

Here, we study confirmable traffic where acknowledgments are required to provide reception guarantees. The results obtained with this study are presented in Fig.~\ref{confirmablestudy} for the same data generation rate and data collection period as was used in the unconfirmable traffic study. In case an acknowledgment is not received by a device in the expected window(s), the device re-transmits the same packet up to 8 times before it is dropped. In the case of Legacy LoRaWAN and Delayed LoRaWAN, the acknowledgment message is an empty message with only a MAC header which confirms reception. In \emph{FREE}, the acknowledgment message includes a bitmap of the same length as the number of slots in the frame, as explained in section~\ref{dcr}.

\begin{figure*}[!ht]
  \centering
  \subfloat[Network Energy Consumption in Unconfirmable Traffic]{
    \includegraphics[width=0.7\columnwidth]{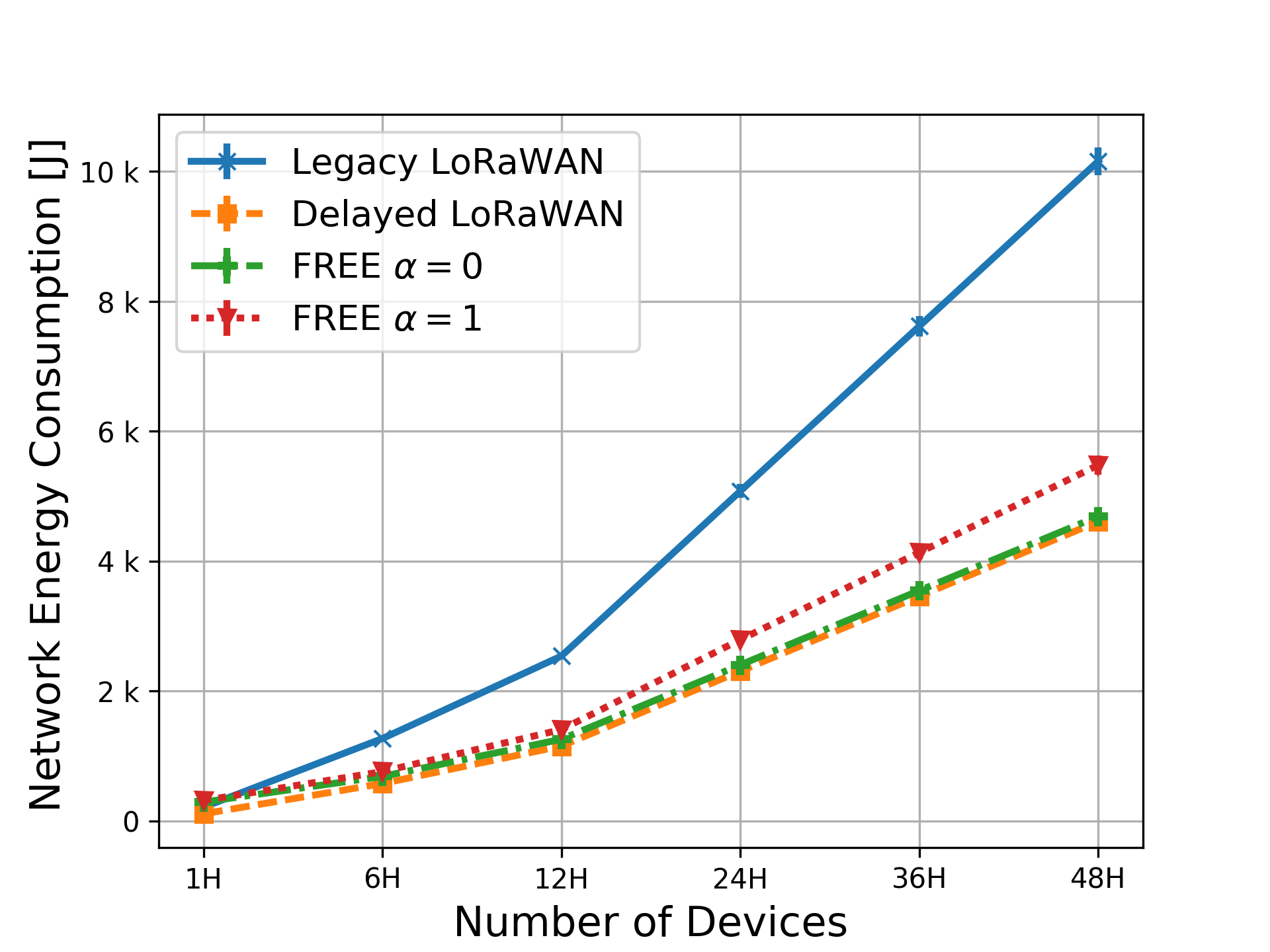}\label{UVenergy}
  }
  \subfloat[Device Lifetime in Unconfirmable Traffic]{
    \includegraphics[width=0.7\columnwidth]{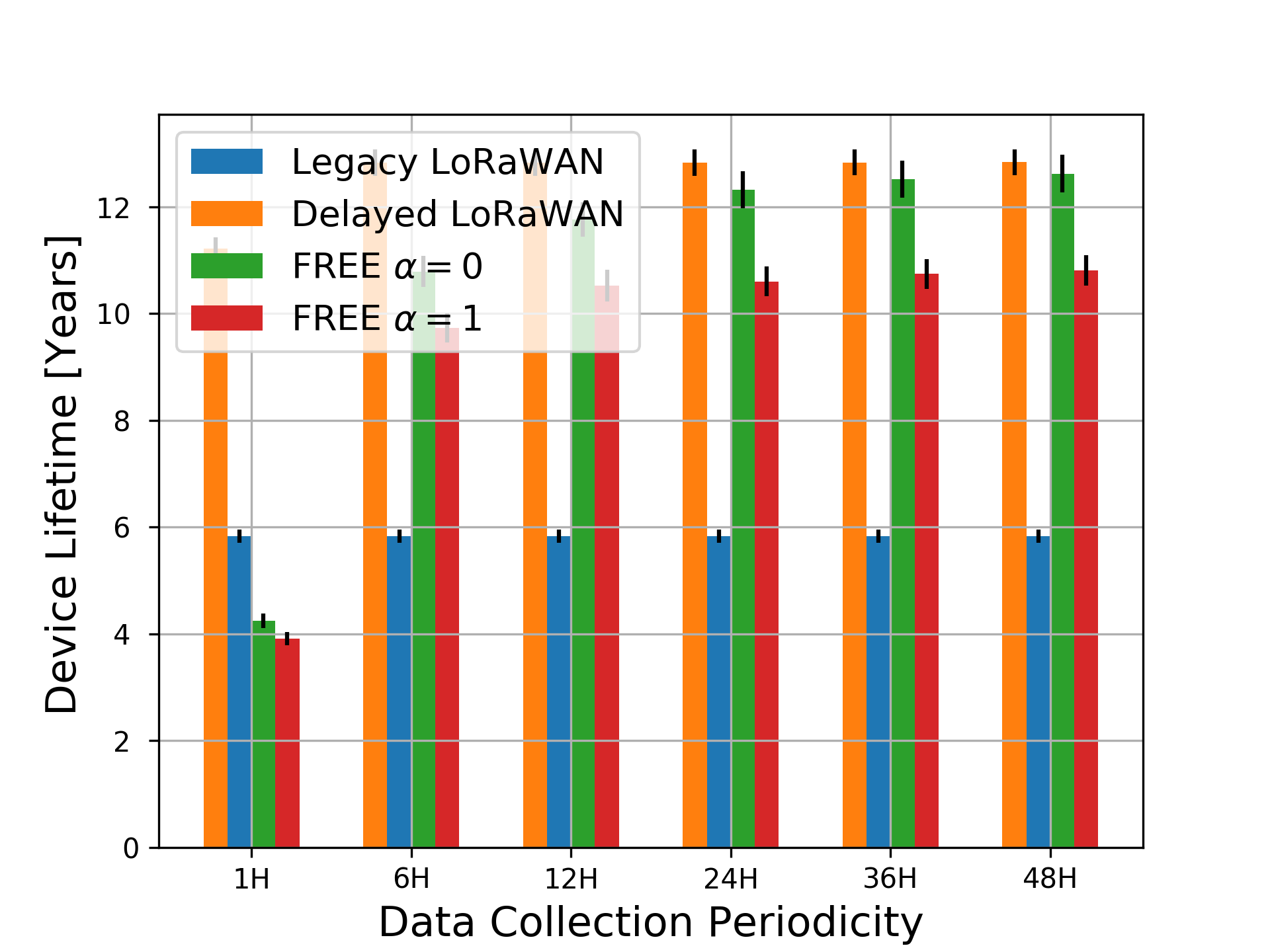}\label{UVlife}
  }  
  \subfloat[Data Delivery Ratio in Unconfirmable Traffic]{
    \includegraphics[width=0.7\columnwidth]{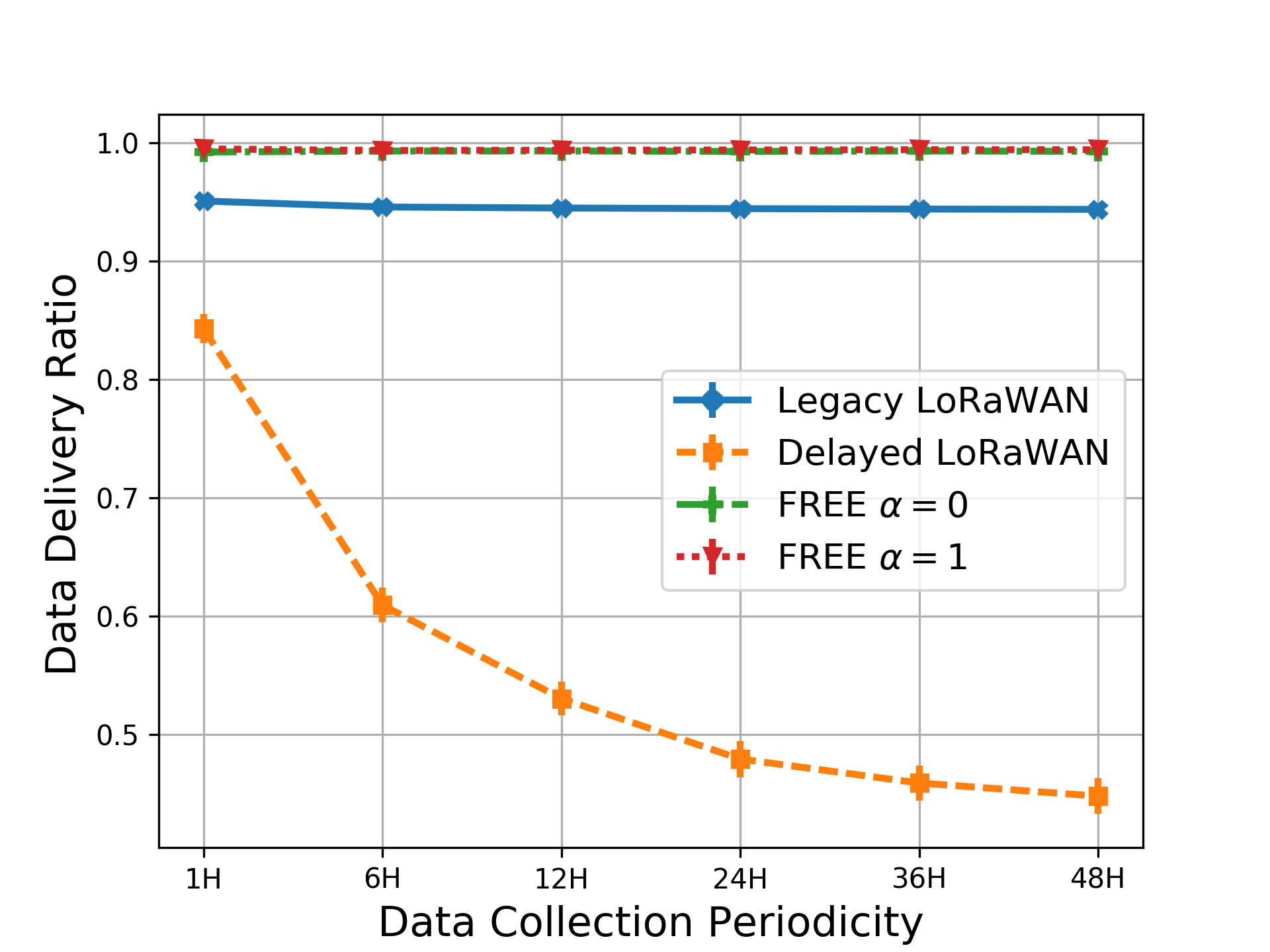}\label{UVder}
  }
  \vspace{-0.35cm}  
  \subfloat[Collection Time in Unconfirmable Traffic]{
    \includegraphics[width=0.7\columnwidth]{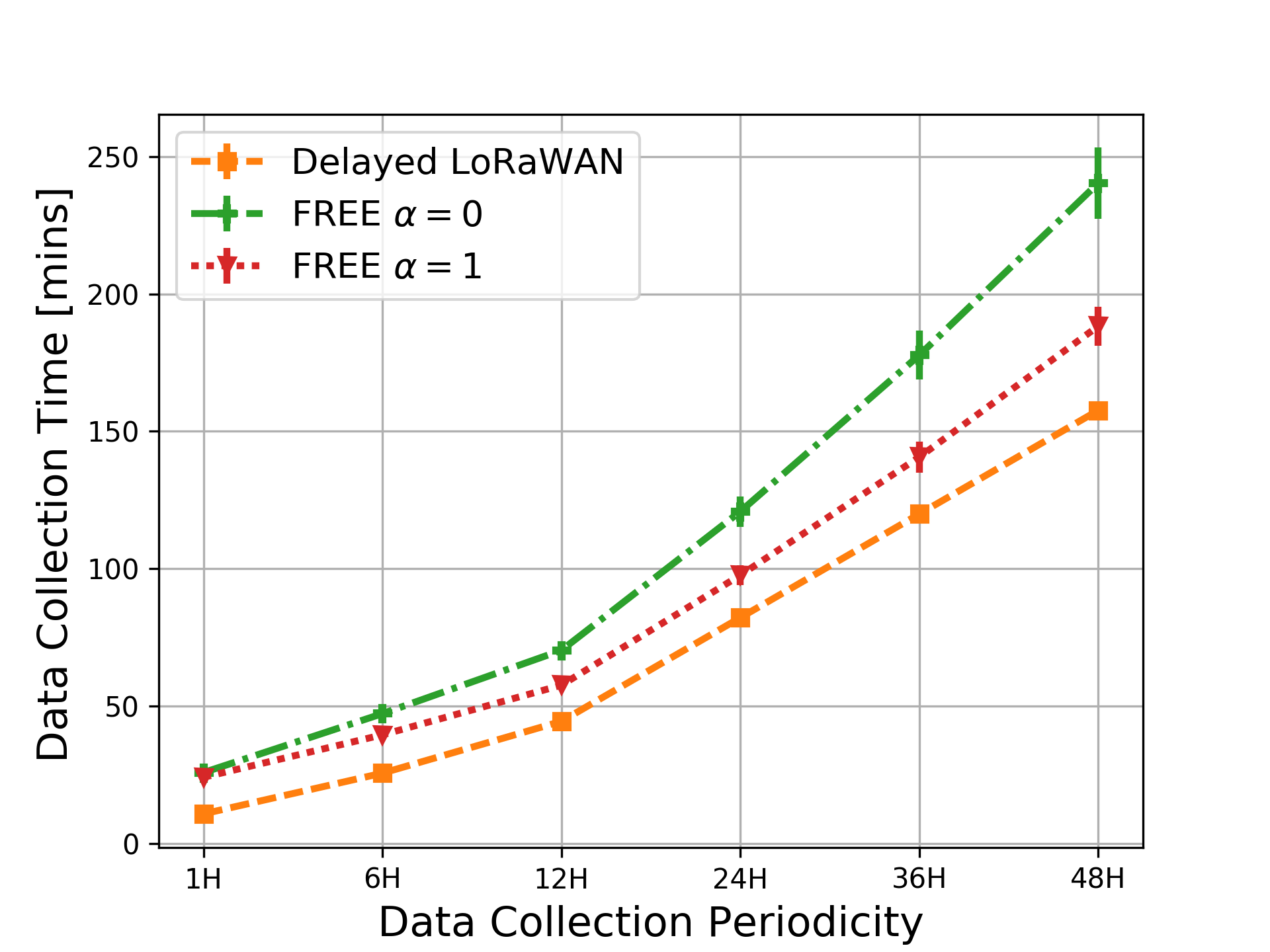}\label{UVtime}
  }
  \subfloat[Network Energy Consumption in Confirmable Traffic]{
    \includegraphics[width=0.7\columnwidth]{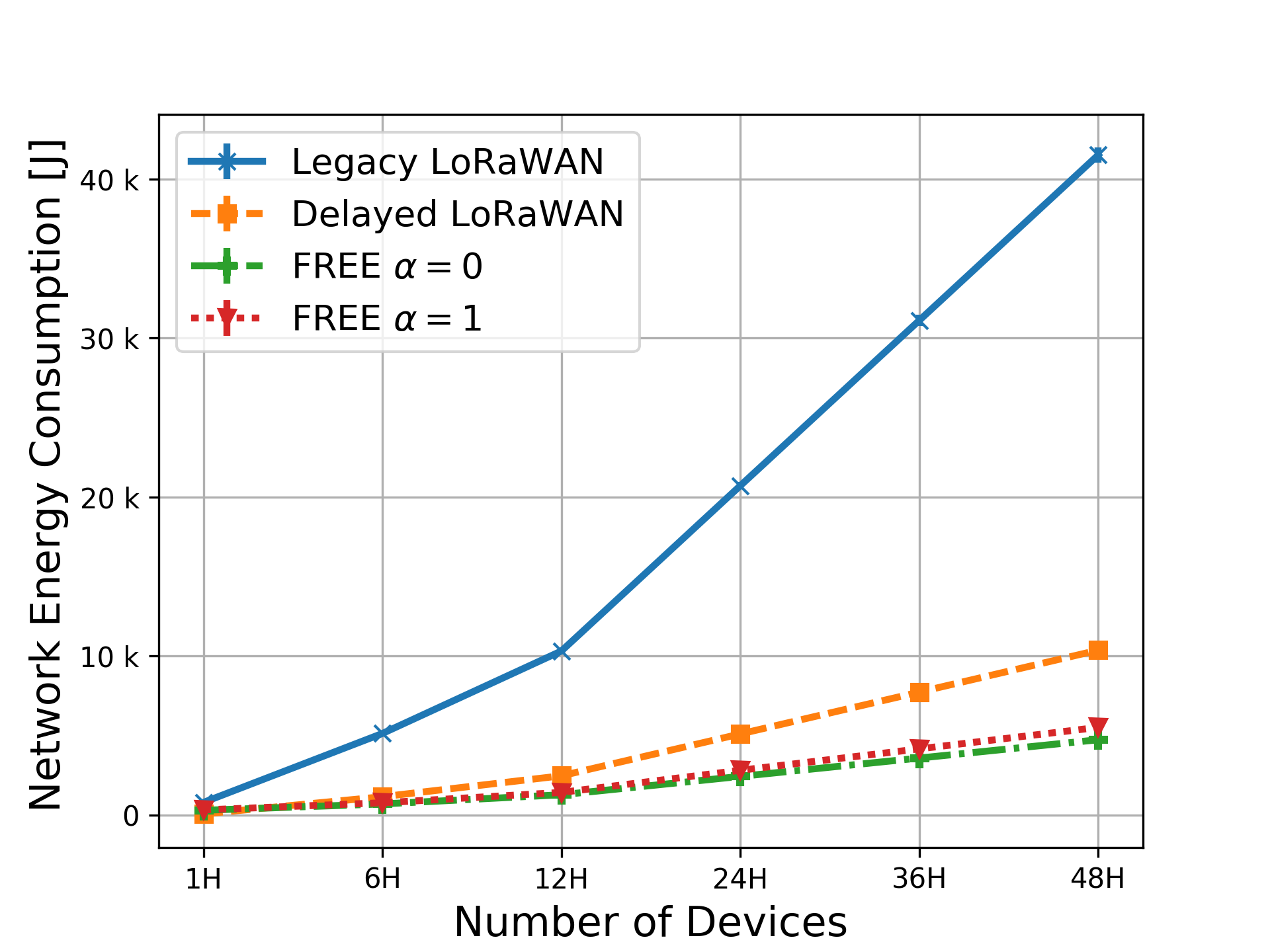}\label{CVenergy}
  }
  \subfloat[Device Lifetime in Confirmable Traffic]{
    \includegraphics[width=0.7\columnwidth]{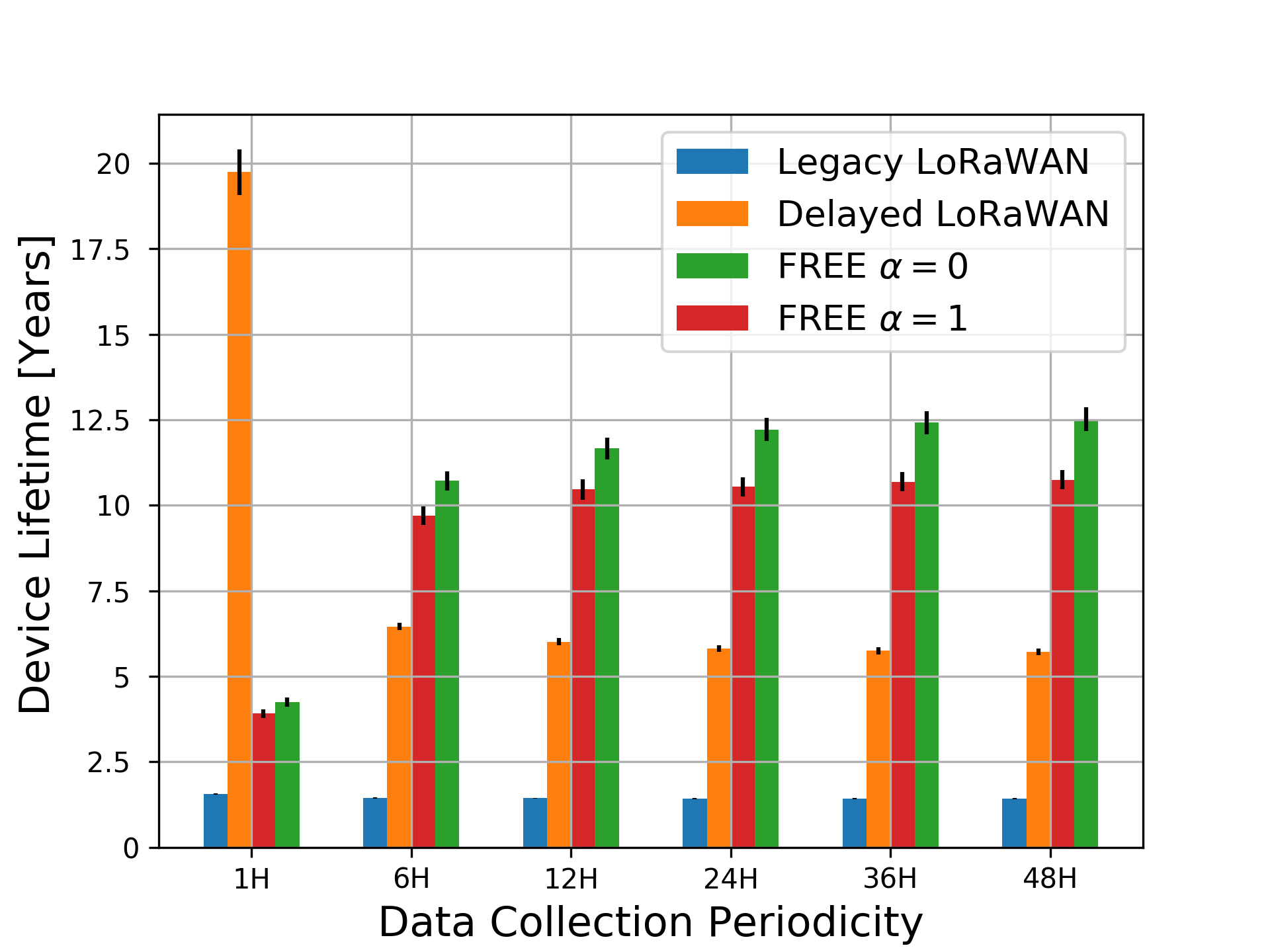}\label{CVlife}
  }
  \vspace{-0.35cm}    
  \subfloat[Data Delivery Ratio in Confirmable Traffic]{
    \includegraphics[width=0.7\columnwidth]{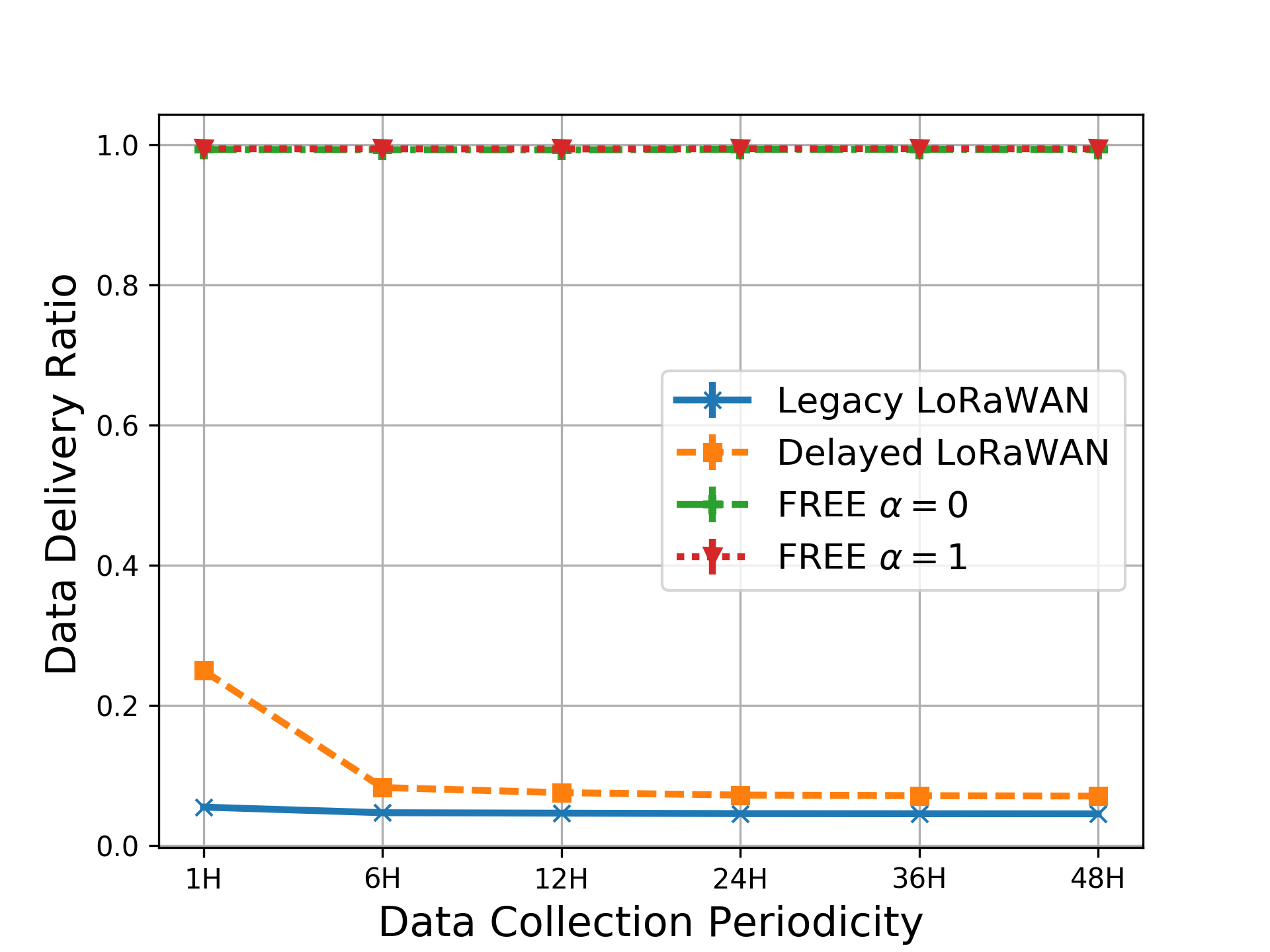}\label{CVder}
  }
  \subfloat[Collection Time in Confirmable Traffic]{
    \includegraphics[width=0.7\columnwidth]{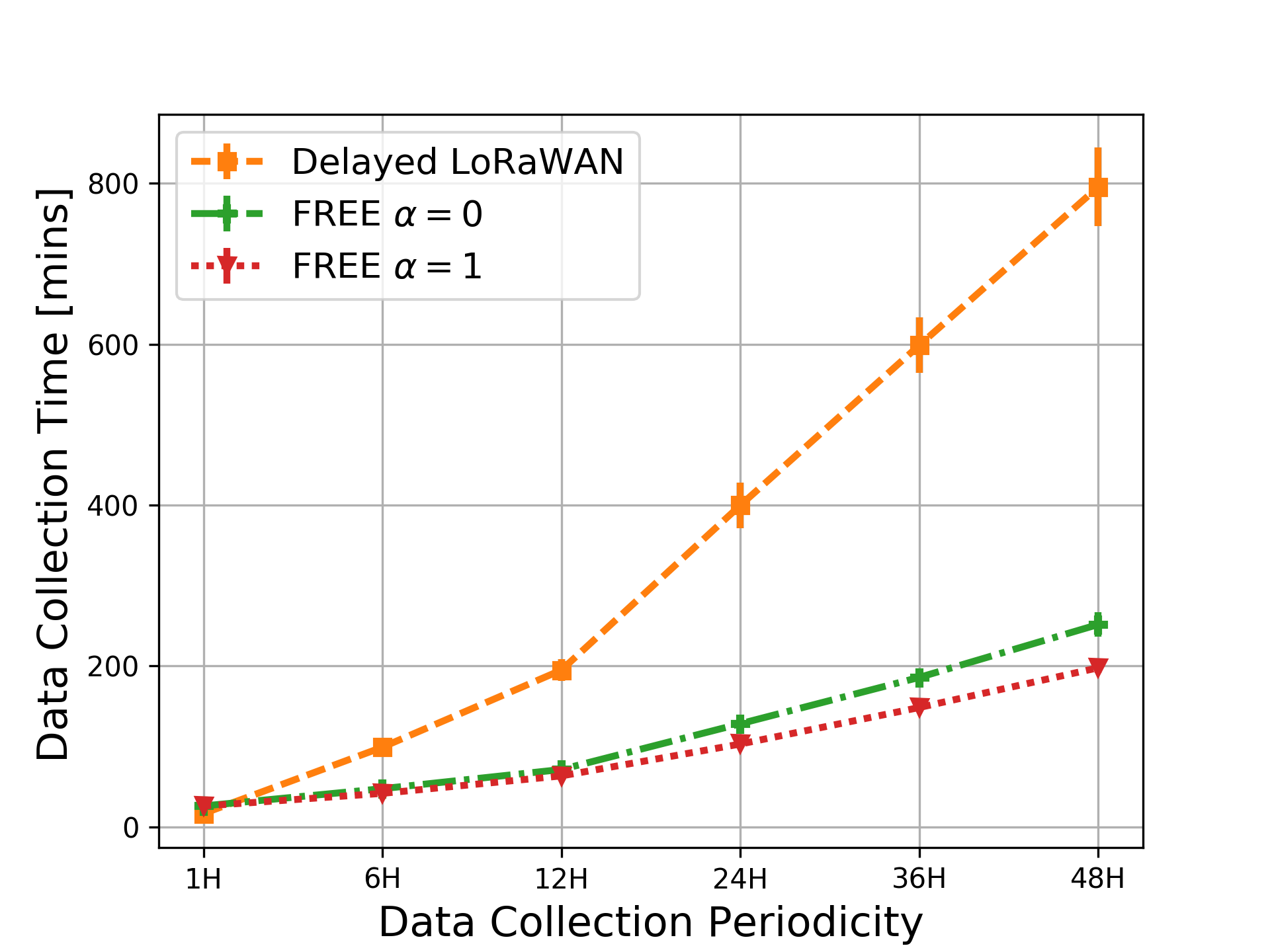}\label{CVtime}
  }
  \caption{Variable Collection Time Periodicity in Unconfirmable and Confirmable Traffic Type}\label{UConVarstudy}
  \vspace{-0.35cm}
\end{figure*}

In terms of energy consumption, Legacy and Delayed LoRaWAN schemes are affected badly when the network size increases with a device's lifetime dropping to less than $2$ years and to about $5$ years, respectively, for a network with $2000$ devices (see Fig.~\ref{Cenergy} and Fig.~\ref{Clife}). This is due to the overhead of retransmissions and confirming reception that the network requires in this traffic case. For instance, a network with $1000$ devices increases the number of transmissions by $7.4$ and $6.4$ times, respectively for Legacy LoRaWAN and Delayed LoRaWAN compared to a unconfirmable traffic (see Fig.~\ref{Utrans} and Fig.~\ref{Ctrans}). In \emph{FREE},  devices can still survive $10$ plus years in both schemes and for all evaluated network sizes. This is because the number of transmissions in this case does not increase much compared to the unconfirmable case due to lack of collisions. A retransmission only takes place due to the loss of an uplink transmission or an acknowledgment because of channel fading. However, these losses happen with low probability. 

In the \emph{FREE} schemes, the increase in the overall data collection time is small compared to the unconfirmable traffic (see Fig.~\ref{Utime} and Fig.~\ref{Ctime}). However, in the case of Delayed LoRaWAN, the difference between the two traffic types is remarkable and gets worse with an increase in network size for the confirmable traffic (see Fig.~\ref{Ctime}) because of the re-transmission overhead. In \emph{FREE}, the spreading factor distribution in this case is identical to the unconfirmable case and this is the reason why the overall data collection time is lower and the overall energy consumption is higher when $\alpha=1$ compared to the case when $\alpha=0$.

In contrast, the confirmable traffic does not improve the overall DDR in case of Legacy and Delayed LoRaWANs. In fact, the overall number of collisions increases (see Fig.~\ref{Ccoll}), which further reduces the overall DDR (see Fig.~\ref{Cder}). However, the collisions are not the main reason for this low DDR, but the gateway duty cycle limitation (see.~\ref{Cnoack}). Fig.~\ref{Cnoack} presents the number of uplink transmissions that the gateway has received but cannot acknowledge in either of the two receive windows due to the duty cycle limitation. The above observation is in line with the analytical analysis in \cite{pop2017does}. In addition to that, packet loss in the uplink (see Fig.~\ref{Clost}) and the downlink (see Fig.~\ref{Cacklost}) due to channel fading also has a negative impact on the DDR, but is not comparable to the impact of the other factors. In Legacy and Delayed LoRaWAN, the DDR drops to less than 20\% for a network with 300 devices and continues to drop with increasing number of devices, showing the well known scalability issue for LoRaWAN with confirmable traffic. It should be noted that for very small network sizes, Delayed LoRaWAN achieves higher network lifetime than the \emph{FREE} solutions without sacrificing the overall DDR (see Fig.~\ref{Clife} and Fig.~\ref{Cder} at 10 devices) due to avoiding the overhead of the joining and synchronization phase.

The \emph{FREE} schemes experience almost no collisions which is due to the synchronization and the transmission power control algorithms. In addition to that, the acknowledgments of a complete frame are grouped in one message, which is sent in the last slot of the frame. In this case, the periodicity of this slot is guaranteed to obey the duty cycle of the channel as the frame length is designed to obey it (see Algorithm~\ref{slandsn}).

\subsection{Collection Periodicity and Delay Elasticity Study} \label{dcps}

Here, we study the impact of varying the data collection periodicity on the overall energy and collection time, which depends on the delay elasticity of the buffered data. The results are gathered from a network with $1000$ devices and we vary the periodicity from $1-hour$ to $48-hours$. The results are presented in Fig.~\ref{UConVarstudy} for unconfirmable and confirmable traffic types, respectively.

For the unconfirmable traffic type, the \emph{FREE} schemes are more energy efficient than Legacy LoRaWAN until data collections are performed as frequently as once per hour. In this case, Legacy LoRaWAN becomes more energy efficient as it avoids the overhead of the joining and synchronization phase without sacrificing the overall DDR (see Fig.~\ref{UVder}). Although Delayed LoRaWAN is the most energy efficient and fastest scheme for unconfirmable traffic, it comes at the expense of the overall DDR. The trend for the confirmable traffic type is consistent; the \emph{FREE} schemes surpass Legacy and Delayed LoRaWANs in terms of  device lifetime (see Fig.~\ref{CVlife}) and data collection time (see Fig.~\ref{CVtime}). This happens without sacrificing the overall DDR (see Fig.~\ref{CVder}) for all the presented periodicities. It should be clear that at this network size, Legacy and Delayed LoRaWANs deliver almost no data, which again highlights the scalability problem of Aloha-based systems for the confirmable traffic type.

Generally speaking, increasing the collection time periodicity minimizes the impact of joining and synchronization overhead of the \emph{FREE} schemes, which increases the overall device lifetime. However, this also increases the overall collection time as devices have more data to transmit. The opposite holds true in case of collecting the data more frequently. The aforementioned statements are valid for both traffic types and the trend can be seen in (Fig.~\ref{UVlife} and Fig.~\ref{UVtime}) for unconfirmable traffic and in (Fig.~\ref{CVenergy} and Fig.~\ref{CVtime}) for confirmable traffic. Therefore, a balance is required to achieve both, a reasonable device lifetime and fast data collection. From this study, we consider a data collection periodicity of $12-hours$ to be a good option. For generalization, data collection should be performed when devices have buffered roughly $2.5 KB$ of data. This amount of data represents a good balance between fast data collection and device lifetime.

\section{Related Work} \label{relatedwork}

The only other work that addresses bulk data transmission in LoRaWAN is our earlier work in ~\cite{offline2019}. In \cite{offline2019}, we considered a static network configuration by assuming that the global state information of the network is known before each data collection. In addition to that, the spreading factors are assumed to be fully orthogonal and we have not discussed how the synchronization aspect or the acknowledged transmissions can be handled. By exploiting these assumptions, an \textit{offline} allocation was proposed, which is opposite to the online allocation that we propose herein. Although the offline allocation can achieve somewhat lower energy consumption than \emph{FREE}, it limits the scope of the approach because of its assumptions. In addition to that, it requires a lot of information to be sent by the gateway, which is not realistic due to the duty cycle limitation. In contrast, \emph{FREE} is independent of the network topology and application(s) (i.e. event-based or periodic). Additionally, it realistically handles the scheduling, synchronization and acknowledged transmissions.

Bulk data transmission over wireless channels has been discussed in a variety of contexts. In~\cite{li2009block}, an optimized transport layer for bulk data transmission in 802.11 was proposed. Also of interest is~\cite{Flush} and ~\cite{gummeson2012flit}, which provides a bulk data transmission protocol for 802.15.4 and RFID, respectively. Our work looks at LoRaWAN in contrast to the other technologies; specifically, we focus on reliable and energy-efficient bulk data collection. Supporting bulk data transmissions over LoRaWAN is constrained by various limitations (e.g. duty cycle, multiple non-orthogonal spreading factors, etc.), which is not the case for wireless networks in comparable previous studies.

A part of our contribution is replacing the Aloha-MAC protocol of standard LoRaWAN with a coordinated MAC. This approach has been taken before in the LoRaWAN literature to improve its scalability. In \cite{reynders2018improving}, time is structured into frames and sub-frames to alleviate collisions. The gateway transmits a beacon before every frame to synchronize the transmissions and guide the spreading factor allocation. If a device wants to send a packet, it has to wait for a beacon first. Despite the time frame structure, the access within sub-frames is still based on Aloha, which means this technique cannot eliminate collisions. In \cite{haxhibeqiri2018low}, the synchronization process is initiated by end devices and the gateway centrally calculates the complete schedule of each device. Here, the schedule is calculated based on the application requirements such as data periodicity and is sent back to each device using a Bloom filter to minimize the message length. Due to the Bloom filters probabilistic data structure, multiple devices may share the same slot and, therefore, this technique does not eliminate collisions either. As the gateway consumes a lot of its duty cycle to transmit the schedules to end devices, this scheme also affects scalability. The two aforementioned approaches are studied for unconfirmable traffic type only and therefore, the scalability question is still not addressed for the confirmable traffic type.

A closely related work to ours is presented in~\cite{zorbas2018collision}, where the use of a drone as a gateway in LoRaWAN was suggested for data collection applications. Nevertheless, this work does not enable bulk data transmissions; specifically, end devices transmit their data as it is generated when the gateway is ready. Perfect synchronization between the gateway and end devices is assumed based on a shared global clock. This leads to uncontrolled de-synchronizations and, thus, severe collisions in the long term. In addition to that the duty cycle limitation is not considered in the design and the spreading factors are assumed to be perfectly orthogonal, leading to overly optimistic results. The technique is studied for unconfirmable transmissions only and has the same bottleneck for confirmable traffic type as the standard LoRaWAN.

\section{Conclusions and Future Work} \label{conclusion}

We addressed the bulk data transmission approach over LoRaWAN. This approach is ideal for various data collection applications in LoRaWAN, however, it is hitherto unreported in the literature. For that, we proposed \emph{FREE} as a fine-grained scheduling scheme for reliable and energy-efficient data collection. In order to compute \emph{FREE}'s transmission schedule, an overhead phase is required to manage the allocations and synchronize the network. We found this overhead is marginal and is minimized when the collection periodicity is large. \emph{FREE} schedules concurrent transmissions by using different spreading factors without collisions and by grouping acknowledgments, which are both bottlenecks of the standard LoRaWAN. We simulated and compared \emph{FREE} to two other approaches for unconfirmable and confirmable transmissions. Our results showed that \emph{FREE} scales well, achieved almost 100\% data delivery and over 10 years battery lifetime independent of the transmission type and network size. Comparing to a poor scalability, low data delivery and device lifetime of fewer than $2$ years in case of confirmable traffic type for standard LoRaWAN configurations.

The implementation of \emph{FREE} in a real LoRaWAN to validate \emph{FREE} in practice is our next piece of work. However, there are also other aspects worth exploring, including the use of a listen-before-talk-based MAC to get rid of the duty-cycle regulations. This would expedite the data collection time, but might be at the expense of the energy consumption as the devices have to perform listening, which is known to be a power-consuming technique. Also, the data collection time could be expedited by using multiple gateways. In this case, collaboration between gateways is required to calculate the schedule and disseminate it while still achieving zero collisions. In this work, we focused on minimizing the overall energy consumption, however it would be interesting to examine the scenario of increasing the fairness among devices in terms of the consumed energy. This would require different allocation techniques to exploit the history of the allocations in the previous data collections. This scenario becomes even more interesting when considering the trajectory of a moving gateway (e.g. a gateway on a drone) in case of a data mule use case.



%

\appendices
\section{} \label{annexA}

\begin{figure}[h]
  \centering
  \subfloat[Join-request Message]{
    \includegraphics[width=1\columnwidth]{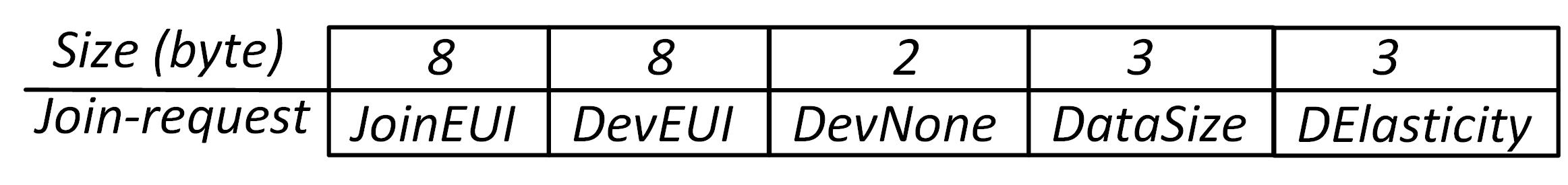}\label{join-request}
  }
  \vspace{-.0ex}
  \subfloat[Join-accept Message]{
    \includegraphics[width=1\columnwidth]{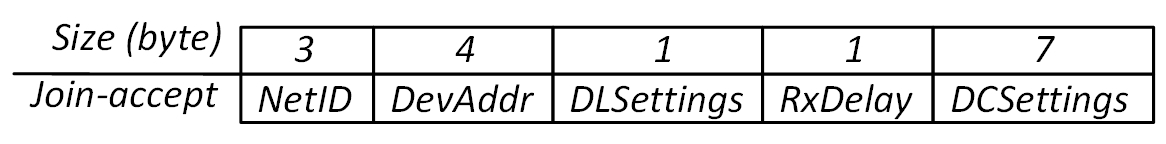}\label{join-accept}
  }
  \vspace{-.0ex}
  \subfloat[DCSettings Field Format]{
    \includegraphics[width=1\columnwidth]{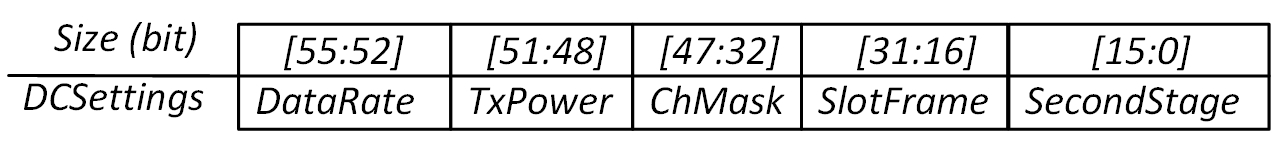}\label{dcsettings}
  }
  \vspace{-.0ex}
  \subfloat[FSettings Message]{
  \includegraphics[width=1\columnwidth]{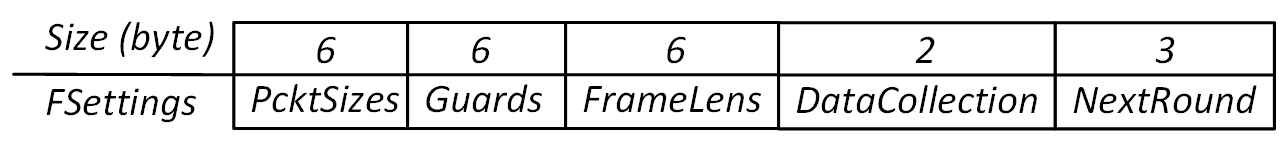}\label{fsettings}
  }
  \caption{Fields of Join-request, Join-accept, and FSettings messages}\label{messages}
  \vspace{-0.35cm}
\end{figure}

Fig.~\ref{messages} shows the message structures of \textit{join-request}, \textit{join-accept}, and \textit{FSettings} messages, which are used in the joining and synchronization phase as illustrated in \S~\ref{synchronizationphase}. The first three fields of the \textit{join-request} message have the same meaning as in a standard LoRaWAN \textit{join-request} message \cite{lorawan2017specs}. Similarly, the first four fields of the \textit{join-accept} message have the same meaning as in a standard LoRaWAN \textit{join-accept} message \cite{lorawan2017specs}. However, the \textit{DCSetting} field is added to let the gateway control the transmission parameters of each device. The format of this field is depicted in Fig.~\ref{dcsettings}, which consists of \textit{DataRate}, \textit{TxPower}, \textit{ChMask}, \textit{SlotFrame}, and \textit{Secondstage} subfields. The first three subfields are used to control the data rate, transmission output power, and the uplink channels and are following the same format as in the LoRaWAN \textit{LinkADRReq} Mac Command \cite{lorawan2017specs}. Finally, \textit{PcktSizes}, \textit{Guards} and \textit{FrameLends} of the \textit{FSettings} message are decoded in a way where the most significant byte in each field corresponds to spreading factor 7, the following byte to spreading factor 8 and so on.


\ifCLASSOPTIONcompsoc
  \section*{Acknowledgments}
\else
  \section*{Acknowledgments}
\fi

This publication has emanated from research conducted with the financial support of Science Foundation Ireland (SFI) and is co-funded under the European Regional Development Fund under Grant Numbers 13/IA/1885 and 13/RC/2077. It has also received funding from the European Union's Horizon 2020 research and innovation program under the Marie Sklodowska-Curie grant agreement No. 713567.

\ifCLASSOPTIONcaptionsoff
  \newpage
\fi



\bibliographystyle{IEEEtran}
\bibliography{biblist}

\begin{thebibliography}{10}
\providecommand{\url}[1]{#1}
\csname url@samestyle\endcsname
\providecommand{\newblock}{\relax}
\providecommand{\bibinfo}[2]{#2}
\providecommand{\BIBentrySTDinterwordspacing}{\spaceskip=0pt\relax}
\providecommand{\BIBentryALTinterwordstretchfactor}{4}
\providecommand{\BIBentryALTinterwordspacing}{\spaceskip=\fontdimen2\font plus
\BIBentryALTinterwordstretchfactor\fontdimen3\font minus
  \fontdimen4\font\relax}
\providecommand{\BIBforeignlanguage}[2]{{%
\expandafter\ifx\csname l@#1\endcsname\relax
\typeout{** WARNING: IEEEtran.bst: No hyphenation pattern has been}%
\typeout{** loaded for the language `#1'. Using the pattern for}%
\typeout{** the default language instead.}%
\else
\language=\csname l@#1\endcsname
\fi
#2}}
\providecommand{\BIBdecl}{\relax}
\BIBdecl

\bibitem{adelantado2017understanding}
F.~Adelantado, X.~Vilajosana, P.~Tuset-Peiro, B.~Martinez, J.~Melia-Segui, and
  T.~Watteyne, ``{Understanding the Limits of LoRaWAN},'' \emph{IEEE
  Communications Magazine}, vol.~55, no.~9, pp. 34--40, 2017.

\bibitem{semtech2015lora}
{Semtech Corporation}, ``{AN1200.22, LoRa™ Modulation Basics},''
  \url{www.semtech.com/uploads/documents/an1200.22.pdf}, 2015, online; accessed
  30-September-2019.

\bibitem{appsurvey}
J.~Haxhibeqiri, E.~De~Poorter, I.~Moerman, and J.~Hoebeke, ``{A Survey of
  LoRaWAN for IoT: From Technology to Application},'' \emph{Sensors}, vol.~18,
  no.~11, 2018.

\bibitem{bor2016lora}
M.~C. Bor, U.~Roedig, T.~Voigt, and J.~M. Alonso, ``{Do LoRa Low-power
  Wide-area Networks Scale?}'' in \emph{Proceedings of the 19th ACM
  International Conference on Modeling, Analysis and Simulation of Wireless and
  Mobile Systems (MSWiM)}.\hskip 1em plus 0.5em minus 0.4em\relax ACM, 2016,
  pp. 59--67.

\bibitem{energymodel}
L.~Casals, B.~Mir, R.~Vidal, and C.~Gomez, ``{Modeling the Energy Performance
  of LoRaWAN},'' \emph{Sensors}, vol.~17, no.~10, 2017.

\bibitem{pop2017does}
A.-I. Pop, U.~Raza, P.~Kulkarni, and M.~Sooriyabandara, ``{Does Bidirectional
  Traffic Do More Harm Than Good in LoRaWAN based LPWA Networks?}'' in
  \emph{IEEE Global Communications Conference (GLOBECOM)}.\hskip 1em plus 0.5em
  minus 0.4em\relax IEEE, 2017, pp. 1--6.

\bibitem{hodge2014wireless}
V.~J. Hodge, S.~O'Keefe, M.~Weeks, and A.~Moulds, ``{Wireless Sensor Networks
  for Condition Monitoring in the Railway Industry: A Survey},'' \emph{IEEE
  Transactions on Intelligent Transportation Systems}, vol.~16, no.~3, pp.
  1088--1106, 2014.

\bibitem{nellore2016survey}
K.~Nellore and G.~Hancke, ``{A Survey on Urban Traffic Management System Using
  Wireless Sensor Networks},'' \emph{Sensors}, vol.~16, no.~2, p. 157, 2016.

\bibitem{airquality}
J.~{Toussaint}, N.~{El Rachkidy}, and A.~{Guitton}, ``{Performance Analysis of
  the On-the-air Activation in LoRaWAN},'' in \emph{2016 IEEE 7th Annual
  Information Technology, Electronics and Mobile Communication Conference
  (IEMCON)}, Oct 2016, pp. 1--7.

\bibitem{smartmetering}
N.~{Varsier} and J.~{Schwoerer}, ``{Capacity Limits of LoRaWAN Technology for
  Smart Metering Applications},'' in \emph{2017 IEEE International Conference
  on Communications (ICC)}, May 2017, pp. 1--6.

\bibitem{reventador}
G.~Werner-Allen, K.~Lorincz, J.~Johnson, J.~Lees, and M.~Welsh, ``{Fidelity and
  Yield in a Avolcano Monitoring Sensor Network},'' in \emph{Proceedings of the
  7th symposium on Operating systems design and implementation}.\hskip 1em plus
  0.5em minus 0.4em\relax USENIX Association, 2006, pp. 381--396.

\bibitem{permafrost}
A.~Hasler, I.~Talzi, J.~Beutel, C.~Tschudin, and S.~Gruber, ``{Wireless Sensor
  Networks in Permafrost Research-concept, Requirements, Implementation and
  Challenges},'' in \emph{Proc. 9th International Conference on Permafrost
  (NICOP 2008)}, vol.~1, 2008, pp. 669--674.

\bibitem{baggio2005wireless}
A.~Baggio, ``{Wireless Sensor Networks in Precision Agriculture},'' in
  \emph{ACM Workshop on Real-World Wireless Sensor Networks (REALWSN 2005),
  Stockholm, Sweden}, vol.~20.\hskip 1em plus 0.5em minus 0.4em\relax ACM,
  2005.

\bibitem{datamuling}
M.~Di~Francesco, S.~K. Das, and G.~Anastasi, ``{Data Collection in Wireless
  Sensor Networks with Mobile Elements: A Survey},'' \emph{ACM Transactions on
  Sensor Networks (TOSN)}, vol.~8, no.~1, p.~7, 2011.

\bibitem{zorbas2018collision}
D.~Zorbas and B.~O’Flynn, ``{Collision-Free Sensor Data Collection using
  LoRaWAN and Drones},'' in \emph{Global Information Infrastructure and
  Networking Symposium (GIIS)}.\hskip 1em plus 0.5em minus 0.4em\relax IEEE,
  Oct 2018.

\bibitem{etsi}
{ETSI ERM TG28}, ``{Electromagnetic Compatibility and Radio Spectrum Matters
  (ERM); Short Range Devices (SRD); Radio equipment to be used in the 25 MHz to
  1 000 MHz Frequency Range with Power Levels Ranging up to 500 mW},'' 2012,
  v2.4.1.

\bibitem{croce2018imperfect}
D.~Croce, M.~Gucciardo, S.~Mangione, G.~Santaromita, and I.~Tinnirello,
  ``{Impact of LoRa Imperfect Orthogonality: Analysis of Link-Level
  Performance},'' \emph{IEEE Communications Letters}, vol.~22, no.~4, pp.
  796--799, April 2018.

\bibitem{demirkol2006mac}
I.~Demirkol, C.~Ersoy, and F.~Alagoz, ``{MAC Protocols for Wireless Sensor
  Networks: A Survey},'' \emph{IEEE Communications Magazine}, vol.~44, no.~4,
  pp. 115--121, 2006.

\bibitem{liang2008koala}
R.~{Musaloiu-E.}, C.~M. {Liang}, and A.~{Terzis}, ``{Koala: Ultra-Low Power
  Data Retrieval in Wireless Sensor Networks},'' in \emph{2008 International
  Conference on Information Processing in Sensor Networks (ipsn 2008)}, April
  2008, pp. 421--432.

\bibitem{abdelfadeel2018fadr}
K.~Q. Abdelfadeel, V.~Cionca, and D.~Pesch, ``{Poster: A Fair Adaptive Data
  Rate Algorithm for LoRaWAN},'' in \emph{International Conference on Embedded
  Wireless Systems and Networks}, ser. EWSN '18.\hskip 1em plus 0.5em minus
  0.4em\relax ACM, 2018.

\bibitem{lora2013calculator}
{Semtech Corporation}, ``{LoRa Modem Design Guide},''
  \url{www.semtech.com/uploads/documents/LoraDesignGuide\_STD.pdf}, 2013,
  online; accessed 30-September-2019.

\bibitem{reynders2016range}
B.~Reynders, W.~Meert, and S.~Pollin, ``{Range and Coexistence Analysis of Long
  Range Unlicensed Communication},'' in \emph{23rd International Conference on
  Telecommunications (ICT)}, May 2016, pp. 1--6.

\bibitem{understanding2012}
E.~Jacobsen, ``{Understanding and Relating Eb/No, SNR, and other Power
  Efficiency Metrics},'' \url{www.dsprelated.com/showarticle/168.php}, 2012,
  accessed: 2018-01-04.

\bibitem{de2005high}
D.~S. De~Couto, D.~Aguayo, J.~Bicket, and R.~Morris, ``{A High-throughput Path
  Metric for Multi-hop Wireless Routing},'' \emph{Wireless networks}, vol.~11,
  no.~4, pp. 419--434, 2005.

\bibitem{offline2019}
D.~Zorbas, K.~Q. Abdelfadeel, V.~Cionca, D.~Pesch, and B.~O'Flynn, ``{Offline
  scheduling algorithms for Time-slotted LoRa-based Bulk Data Transmission},''
  in \emph{IEEE 5th World Forum on Internet of Things (WFIoT)}.\hskip 1em plus
  0.5em minus 0.4em\relax IEEE, 2019, pp. 1--6.

\bibitem{abdelfadeel2018fair}
\vspace{0mm}K. Q.~Abdelfadeel, V.~Cionca, and D.~Pesch, ``{Fair Adaptive Data
  Rate Allocation and Power Control in LoRaWAN},'' in \emph{2018 IEEE 19th
  International Symposium on "A World of Wireless, Mobile and Multimedia
  Networks" (WoWMoM)}, June 2018, pp. 14--15.

\bibitem{li2009block}
M.~Li, D.~Agrawal, D.~Ganesan, A.~Venkataramani, and H.~Agrawal,
  ``{Block-switched Networks: A New Paradigm for Wireless Transport.}'' in
  \emph{NSDI}, vol.~9, 2009, pp. 423--436.

\bibitem{Flush}
S.~Kim, R.~Fonseca, P.~Dutta, A.~Tavakoli, D.~Culler, P.~Levis, S.~Shenker, and
  I.~Stoica, ``{Flush: A Reliable Bulk Transport Protocol for Multihop Wireless
  Networks},'' in \emph{SenSys '07}.\hskip 1em plus 0.5em minus 0.4em\relax
  ACM, 2007, pp. 351--365.

\bibitem{gummeson2012flit}
J.~Gummeson, P.~Zhang, and D.~Ganesan, ``{Flit: A Bulk Transmission Protocol
  for RFID-scale Sensors},'' in \emph{Proceedings of the 10th international
  conference on Mobile systems, applications, and services}.\hskip 1em plus
  0.5em minus 0.4em\relax ACM, 2012, pp. 71--84.

\bibitem{reynders2018improving}
B.~Reynders, Q.~Wang, P.~Tuset-Peiro, X.~Vilajosana, and S.~Pollin,
  ``{Improving Reliability and Scalability of LoRaWANs through Lightweight
  Scheduling},'' \emph{IEEE Internet of Things Journal}, vol.~5, no.~3, pp.
  1830--1842, 2018.

\bibitem{haxhibeqiri2018low}
J.~Haxhibeqiri, I.~Moerman, and J.~Hoebeke, ``{Low Overhead Scheduling of LoRa
  Transmissions for Improved Scalability},'' \emph{IEEE Internet of Things
  Journal}, vol.~6, no.~2, pp. 3097--3109, 2018.

\bibitem{lorawan2017specs}
{LoRa Alliance, Technical Committee}, ``{LoRaWAN™ 1.1 Specification},''
  \url{lora-alliance.org/sites/default/files/2018-04/lorawantm\_specification\_-v1.1.pdf},
  2017, online; accessed 30-September-2019.

\end{thebibliography}
%


%

\begin{IEEEbiographynophoto}{Khaled Abdelfadeel}
received his MSc and BSc degrees in Electronics and Communications Engineering from Cairo University, Egypt, in 2016 and 2011, respectively. He is currently a PhD student at the School of Computer Science and IT, University College Cork in Ireland. Khaled's research focuses on the resource management and IP interoperability for low-power wide-area wireless networks.
\end{IEEEbiographynophoto}

\begin{IEEEbiographynophoto}{Dimitrios Zorbas}
holds a PhD in Computer Science from the University of Piraeus in Greece. He has worked as post doctoral researcher at Inria Lille -- Nord Europe and at University of La Rochelle in France. He is currently a researcher with the Tyndall National Institute after receiving a Marie Curie fellowship. He is author of more than 40 peer-reviewed publications in the area of computer communications, energy efficiency in networks, and secure communications. He has also worked in several national as well as FP7 and H2020 projects. He is member of the IEEE.
\end{IEEEbiographynophoto}


\begin{IEEEbiographynophoto}{Victor Cionca} is a Senior Researcher with the Nimbus Research Centre at Cork Institute of Technology, Cork, Ireland. He holds a PhD in Wireless Sensor Networks from the University of Limerick. He is conducting research in modelling, controlling, and optimising wireless network behaviour. Research interests include virtualised, multi-owner, wireless meshes, LPWAN, and energy harvesting algorithms.
\end{IEEEbiographynophoto}


\begin{IEEEbiographynophoto}{Dirk Pesch} (S`96-M`00-SM`17) is Professor of Computer Science at University College Cork in Ireland. Prior to joining UCC, Dirk was Professor and Head of Centre of the Nimbus Research Centre at Cork Institute of Technology. Dirk’s research interests focus on architecture, design, algorithms, and performance evaluation of low power, dense and moving wireless/mobile networks and services for Internet of Things and Cyber-Physical System’s applications and interoperability issues associated with IoT applications. He has over 25 years research and development experience in both industry and academia and has co-authored over 200 scientific articles. He is a principle investigator in Science Foundation Ireland funded initiatives such as the CONNECT Centre for Future Networks, the ENABLE research programme on smart communities and the CONFIRM Centre for Smart Manufacturing. He has also been active in many EU funded research projects, including as coordinator, and was involved with two startups. He is on the editorial board a number of international journals, and contributes to international conference organization in his area of expertise. Dirk received a Dipl.Ing. degree from RWTH Aachen University, Germany, and a PhD from the University of Strathclyde, Glasgow, Scotland, both in Electrical \& Electronic Engineering.
\end{IEEEbiographynophoto}



\end{document}